\documentclass[prb,aps,preprint,showpacs,superscriptaddress,groupedaddress,amsmath,amssymb]{revtex4-1} 

\usepackage{graphicx}
\usepackage{threeparttable}
\usepackage{here}
\usepackage{ulem}
\usepackage{color}

\newcommand{\ket}[1] {\left| #1 \right\rangle}

\begin{document}
\title{Superconducting qubit-oscillator circuit beyond the ultrastrong-coupling regime}

\author{Fumiki Yoshihara}
\email{fumiki@nict.go.jp}
\thanks{These authors contributed equally to this work.}
\affiliation{National Institute of Information and Communications Technology, 4-2-1, Nukuikitamachi, 
Koganei, Tokyo 184-8795, Japan}

\author{Tomoko Fuse}
\email{tfuse@nict.go.jp}
\thanks{These authors contributed equally to this work.}
\affiliation{National Institute of Information and Communications Technology, 4-2-1, Nukuikitamachi, 
Koganei, Tokyo 184-8795, Japan}

\author{Sahel Ashhab}
\email{sashhab@qf.org.qa}
\affiliation{Qatar Environment and Energy Research Institute,  
Hamad Bin Khalifa University, Qatar Foundation, 
Doha, Qatar}

\author{Kosuke Kakuyanagi}
\affiliation{NTT Basic Research Laboratories, NTT Corporation, 3-1 Morinosato-Wakamiya, 
Atsugi, Kanagawa 243-0198, Japan}

\author{Shiro Saito}
\affiliation{NTT Basic Research Laboratories, NTT Corporation, 3-1 Morinosato-Wakamiya, 
Atsugi, Kanagawa 243-0198, Japan}

\author{Kouichi Semba}
\email{semba@nict.go.jp}
\affiliation{National Institute of Information and Communications Technology, 4-2-1, Nukuikitamachi, 
Koganei, Tokyo 184-8795, Japan}

\begin{abstract}
The interaction between an atom and the electromagnetic field inside a cavity\cite{Wineland81,Haroche93RMP,Brune96,Raimond01,Mabuchi02,walls2007quantum}
has played a crucial role in the historical development of our understanding of light-matter interaction
and is a central part of various quantum technologies,
such as lasers and many quantum computing architectures.
The emergence of superconducting qubits\cite{Nakamura99Nature,Clarke08Nature}
has allowed the realization of strong\cite{Chiorescu04,Wallraff04} and ultrastrong \cite{DGS07,Niemczyk10,FornDiaz10} coupling between artificial atoms and cavities.
If the coupling strength $g$ becomes as large as the atomic and cavity frequencies
($\Delta$ and $\omega_{\rm o}$ respectively),
the energy
eigenstates including the ground state are predicted to be highly entangled\cite{HeppLieb73}.
This qualitatively new regime can be called the deep strong-coupling regime\cite{Casanova10},
and there has been an ongoing debate\cite{Rzazewski75PRL,Nataf10NatCom,Viehmann11PRL} over
whether it is fundamentally possible to realize this regime in realistic physical systems.
By inductively coupling a flux qubit and an LC oscillator via Josephson junctions,
we have realized circuits with $g/\omega_{\rm o}$ ranging from 0.72 to 1.34 and $g/\Delta\gg 1$.
Using spectroscopy measurements,
we have observed unconventional transition spectra, with patterns resembling masquerade masks,
that are characteristic of this new regime.
Our results provide a basis for ground-state-based entangled-pair generation
and open a new direction of research on strongly correlated light-matter states in circuit-quantum electrodynamics.
\end{abstract}

\maketitle
\begin{figure}
\includegraphics{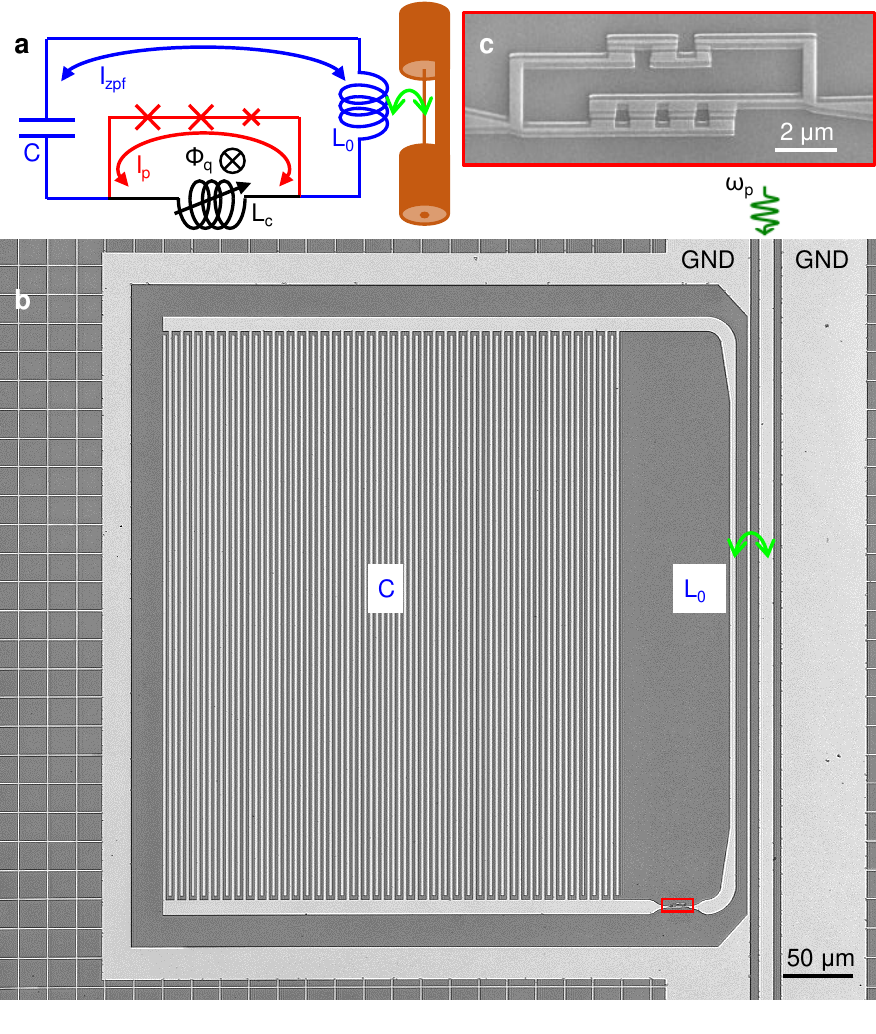}
\caption{\textbf{Superconducting qubit-oscillator circuit.} \textbf{a},~Circuit diagram.
A superconducting flux qubit (red and black) and a superconducting LC oscillator (blue and black) are inductively coupled to each other by sharing a tunable inductance (black).
\textbf{b},~Laser microscope image of the lumped-element LC oscillator inductively coupled to a coplanar transmission line.
\textbf{c},~Scanning electron microscope image of the qubit and the coupler junctions located at the red rectangle in image \textbf{b}.
The coupler, consisting of four parallel Josephson junctions, is tunable via the magnetic flux bias through its loops (see Supplementary Information, sections~S1 and S2, and Fig.~S1).
}
\label{SEM}
\end{figure}

We begin by describing the Hamiltonian of each component in the qubit-oscillator circuit,
which comprises a superconducting flux qubit and an LC oscillator inductively coupled to each other by sharing a tunable inductance $L_{\rm c}$,
as shown in the circuit diagram in Fig.~\ref{SEM}\textbf{a}.

%Flux qubit
The Hamiltonian of the flux qubit can be written in the basis of two states with persistent currents flowing in opposite directions around the qubit loop~\cite{Mooij99Sci},
$|\textrm{L}\rangle_{\rm q}$ and $|\textrm{R}\rangle_{\rm q}$, as $\mathcal{H}_{\rm q} = -\hbar(\Delta \sigma_x + \varepsilon \sigma_z)/2$,
where $\hbar\Delta$ and 
$\hbar \varepsilon=2I_{\rm p}\Phi_0(n_{\phi \rm q}-n_{\phi \rm q0})$
are the tunnel splitting and the energy bias between $|\textrm{L}\rangle_{\rm q}$ and
$|\textrm{R}\rangle_{\rm q}$,
$I_{\rm p}$ is the maximum persistent current,
and $\sigma_{x,\,z}$ are Pauli matrices.
Here, $n_{\phi \rm q}$ is the normalized flux bias through the qubit loop in units of the superconducting flux quantum, $\Phi_0=h/2e$,
and $n_{\phi \rm q0} = 0.5 + k_{\rm q}$,
where $k_{\rm q}$ is the integer that minimizes $|n_{\phi \rm q}-n_{\phi \rm q0}|$.
The macroscopic nature of the persistent-current states enables
strong coupling to other circuit elements.
Another important feature of the flux qubit is its strong anharmonicity:
the two lowest energy levels are well isolated from the higher levels.

%LC oscillator
The Hamiltonian of the LC oscillator can be written as $\mathcal{H}_{\rm o} = \hbar \omega_{\rm o}(\hat{a}^{\dagger}\hat{a} + 1/2)$,
where $\omega_{\rm o} = 1/\sqrt{(L_0+L_{\rm qc})C}$ is the resonance frequency,
$L_{\rm 0}$ is the inductance of the superconducting lead,
$L_{\rm qc}(\simeq L_{\rm c})$ is the inductance across the qubit and coupler (see Supplementary Information, section~S2),
$C$ is the capacitance,
and $\hat{a}$ $(\hat{a}^{\dagger})$ is the oscillator's annihilation (creation) operator.
Figure~\ref{SEM}\textbf{b} shows a laser microscope image of the lumped-element LC oscillator,
where $L_0$ is designed to be as small as possible to maximize the zero-point fluctuations in the current $I_{\rm zpf}=\sqrt{\hbar \omega_{\rm o}/2(L_0 + L_{\rm qc})}$ and hence achieve strong coupling to the flux qubit, while $C$ is adjusted so as to achieve a desired value of $\omega_{\rm o}$.
The freedom of choosing $L_0$ for large $I_{\rm zpf}$ is one of the advantages of lumped-element LC oscillators over coplanar-waveguide resonators for our experiment.
Another advantage is that a lumped-element LC oscillator has only one resonant mode.
Together with the strong anharmonicity of the flux qubit,
we can expect that our circuit will realize the Rabi model~\cite{Rabi37PR,Shimoda56PR,Jaynes63I3E,Braak11PRL},
which is one of the simplest possible quantum models of qubit-oscillator systems,
with no additional energy levels in the range of interest.

%coupler inductance.
The coupling Hamiltonian can be written as\cite{Chiorescu04} $\mathcal{H}_{\rm c}=\hbar g \sigma_z (\hat{a} + \hat{a}^{\dagger})$,
where $\hbar g=MI_{\rm p}I_{\rm zpf}$ is the coupling energy and
$M (\simeq L_{\rm c})$ is the mutual inductance between the qubit and the LC oscillator.
Importantly, a Josephson-junction circuit is used as a large inductive coupler\cite{Bourassa09PRA} (Fig.~\ref{SEM}\textbf{c}), which together with the large $I_{\rm p}$ and $I_{\rm zpf}$ allows us to achieve deep strong coupling.

%total Hamiltonian
The total Hamiltonian of the circuit is then given by
\begin{eqnarray}
\label{Htotal}
\mathcal{H}_{\rm total} = -\frac{\hbar}{2}(\Delta \sigma_x + \varepsilon \sigma_z) + \hbar \omega_{\rm o}(\hat{a}^{\dagger}\hat{a} + \frac{1}{2}) + \hbar g\sigma_z(\hat{a} + \hat{a}^{\dagger}).
\end{eqnarray}
Nonlinearities in the coupler circuit lead to higher-order terms in $(\hat{a} + \hat{a}^{\dagger})$.
The leading-order term can be written as $C_{\rm A2}\hbar g(\hat{a} + \hat{a}^{\dagger})^2$ and is known as the $A^2$ term~\cite{Rzazewski75PRL} in atomic physics.
Since this $A^2$ term can be eliminated from $\mathcal{H}_{\rm total}$ by a variable transformation (see Methods),
we do not explicitly keep it and instead use Eq.~(\ref{Htotal}) for our data analysis.

%Fig2
\begin{figure}
\includegraphics{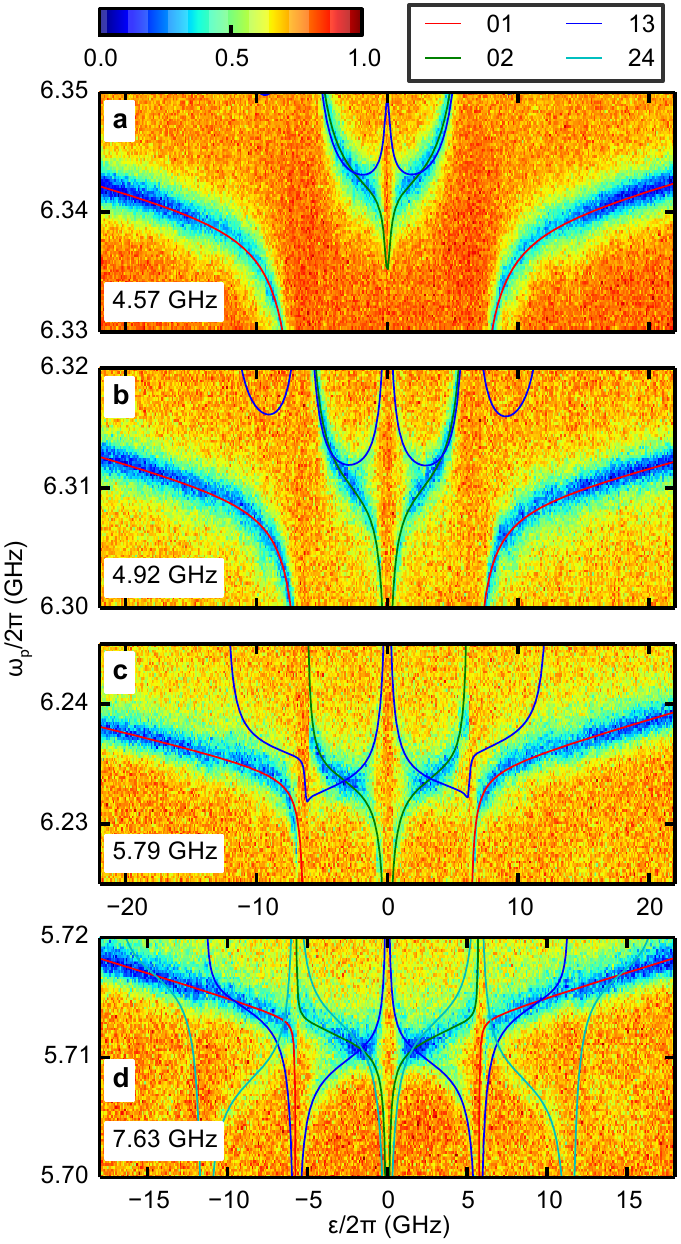}
\caption{\textbf{Transmission spectra for circuits~I and II.}
Calculated transition frequencies $\omega_{ij}^{\rm cal}$ are superimposed
on the experimental results.
As summarized in Table~\ref{params},
panel~\textbf{a} shows
data from circuit~I at $n_{\phi \rm q}= -0.5$,
panel \textbf{b} shows
data from circuit~I at $n_{\phi \rm q}= -1.5$,
panel~\textbf{c} shows
data from circuit~I at $n_{\phi \rm q}= 2.5$,
and panel~\textbf{d} shows
data from circuit~II at $n_{\phi \rm q}= -0.5$.
The values of $g/2\pi$ are written in the panels.
The red, green, blue, and cyan lines indicate the transitions $|0\rangle \to |1\rangle$,
$|0\rangle \to |2\rangle$, $|1\rangle \to |3\rangle$, and $|2\rangle \to |4\rangle$, respectively.
}
\label{PFB}
\end{figure}

%Table 1
\begin{table}
\begin{threeparttable}
\caption{\textbf{Set of parameters obtained from fitting spectroscopy measurements.}}
\begin{tabular}{c c @{\hspace{0.5cm}} c @{\hspace{0.5cm}} c c c c c c}
    \hline
    \hline
     circuit & $n_{\phi \rm q}$ & Figure & $\Delta/2\pi$ & $\omega_{\rm o}/2\pi$ & $g/2\pi$ & $\alpha=g/\omega_{\rm o}$ & $2g/\sqrt{\omega_{\rm o}\Delta}$\\
     &&&(GHz)&(GHz)&(GHz)&&&\\
    \hline
    I & $-0.5$ & 2\textbf{a} & 0.505 & 6.336 & 4.57 & 0.72 & 5.1\\
    I & $-1.5$ & 2\textbf{b}, 3\textbf{a} & 0.430 & 6.306 & 4.92 & 0.78 & 6.0\\
    I & 2.5 & 2\textbf{c} & 0.299 & 6.233 & 5.79 & 0.93 & 8.5\\
    II & $-0.5$ & 2\textbf{d} & 0.441 & 5.711 &7.63 & 1.34 & 9.6\\
    III & $0.5$ & SI6 & 3.84 & 5.588 & 5.63 & 1.01 & 2.4\\

    \hline
    \hline
\end{tabular}
\label{params}
{\small
\begin{tablenotes}
\item[]The parameters are obtained from five sets of spectroscopy data in three circuits.
The column ``Figure'' shows the corresponding figures.
``SI'' stands for Supplementary Information.
\end{tablenotes}
}
\end{threeparttable}
\end{table}

%measurement
Spectroscopy was performed by measuring the transmission spectrum through a coplanar transmission line that is inductively coupled to the LC oscillator (see Supplementary Information, section~S3).
For a systematic study of the $g$ dependence,
five flux bias points in three circuits were used.
Circuit~II is designed to have larger values of $g$ than the other two,
and circuits~I and II are designed to have smaller values of $\Delta$ than circuit~III.
Figures~\ref{PFB}\textbf{a}--\textbf{d} show normalized amplitudes of the transmission spectra $|S_{21}(\omega_{\rm p})|/|S_{21}(\omega_{\rm p})|_{\rm max}$ from circuits~I and II as functions of the flux bias $\varepsilon$ and probe frequency $\omega_{\rm p}$
(see also Supplementary Information, Figs.~S5\textbf{a}-\textbf{d}).
Characteristic patterns resembling masquerade masks can be seen around $\varepsilon = 0$.
At each value of $\varepsilon$, the spectroscopy data was fitted with Lorentzians to obtain the frequencies $\omega_{ij}$ of the transitions
$|i\rangle \to |j\rangle$, where the indices $i$ and $j$ label the energy eigenstates
according to their order in the energy-level ladder, with the index 0 denoting the ground state.
Theoretical fits to $\omega_{ij}$ were obtained by diagonalizing $\mathcal{H}_{\rm total}$,
treating $\Delta$, $\omega_{\rm o}$, and $g$ as fitting parameters.
The obtained parameters are shown in Table~\ref{params}.
The calculated transition frequencies $\omega_{ij}^{\rm cal}$ are superimposed on the measured transmission spectra.
As $g$ increases, the anticrossing gap between the qubit and the oscillator frequencies at $\varepsilon \simeq \pm \omega_{\rm o}$
becomes smaller and the signal from the $|1\rangle \to |3\rangle$ transition gradually transforms from a W shape to a $\Lambda$ shape in the range $|\varepsilon|\lesssim \omega_{\rm o}$.
These features are seen in both the experimental data and the theoretical calculations,
with good agreement between the data and the calculations.
Note that $\omega_{\rm o}$ depends on the qubit state and $\varepsilon$ via $L_{\rm qc}$,
which results in the broad V shape seen in the spectra (see Supplementary Information, section~S2).

%03 and 12 transitions
To capture signals from more transitions,
the transmission spectra in a wider $\omega_{\rm p}$ range and a smaller $\varepsilon$ range were measured, as shown in Fig.~3\textbf{a} for circuit~I at $n_{\phi \rm q} = -1.5$.
As we approach the symmetry point $\varepsilon = 0$, the signals from the $|0\rangle \to |2\rangle$ and $|1\rangle \to |3\rangle$ transitions disappear while signals from the $|0\rangle \to |3\rangle$ and $|1\rangle \to |2\rangle$ transitions appear near $\omega_{03}^{\rm cal}$ and $\omega_{12}^{\rm cal}$.
The appearance and disappearance of the signals are well explained by the transition matrix elements $T_{ij}=\langle i|(\hat{a} + \hat{a}^{\dagger})|j\rangle$ shown in Fig.~3\textbf{b}:
when $\varepsilon \to 0$, $|T_{02}| = |T_{13}| \to 0$
(forbidden transitions),
while $|T_{03}|$ and $|T_{12}|$ are maximum (allowed transitions).
As can be seen from the expression for $T_{ij}$,
these features are directly related to the form of the energy eigenstates and can therefore serve as indicators of the symmetry properties of the energy eigenstates,
similarly to how atomic forbidden transitions are related to the symmetry of atomic wave functions.
%weakness of the signals
The weakness of the signals from the $|0\rangle \to |3\rangle$ and $|1\rangle \to |2\rangle$ transitions
is probably due to dephasing caused by flux fluctuations.
No signals from the $|0\rangle \to |3\rangle$ and $|1\rangle \to |2\rangle$ transitions were observed in circuit~I at $n_{\phi \rm q}= 2.5$ and in circuit~II.
The broad dips at $\omega_{\rm p}/2\pi = 6.2$, 6.38, and 6.45~GHz are the result of a background frequency dependence of the transmission line's transmission amplitude,
and these features can be ignored here.
The feature at 6.2~GHz also contains a narrow signal from another qubit-oscillator circuit that is coupled to the transmission line (see Supplementary Information, section~S3).

%fig 3
\begin{figure}
\includegraphics{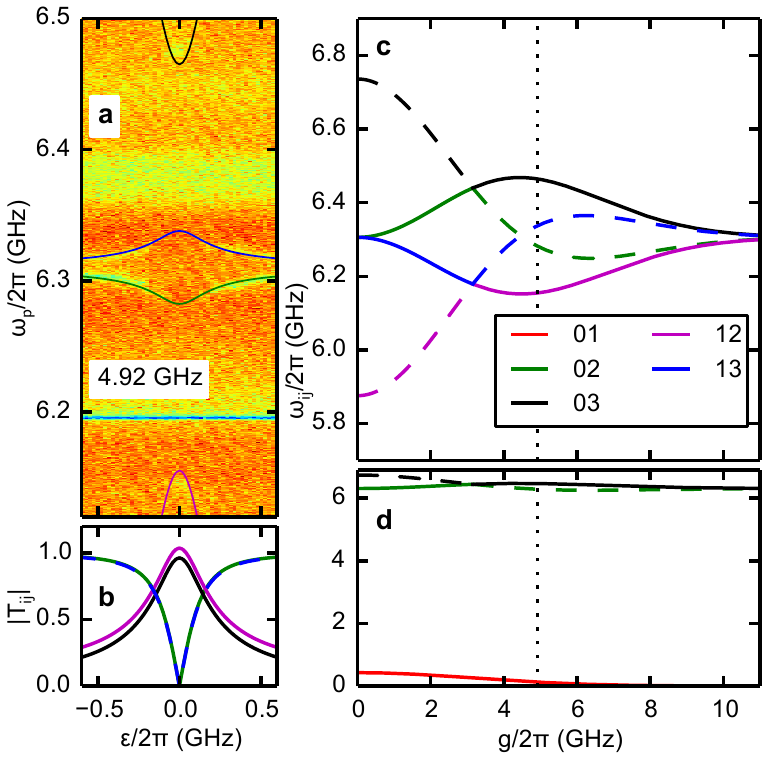}
\caption{\textbf{Selection rules and transmission spectrum around the symmetry point.}
\textbf{a}~Transmission spectrum for circuit~I at $n_{\phi \rm q} = -1.5$ plotted as a function of flux bias $\varepsilon$.
The transition frequencies $\omega_{ij}^{\rm cal}$ superimposed on the experimental result in \textbf{a} and the matrix elements $|T_{ij}|$ plotted in \textbf{b} are calculated using the parameters shown in Table~I, i.e. $\Delta/2\pi =0.430$~GHz, $\omega_{\rm o}/2\pi =6.306$~GHz,
and $g/2\pi =4.92$~GHz.
\textbf{c} The calculated transition frequencies around $\omega_{\rm o}$ and
\textbf{d} from the ground state are plotted as functions of $g$ at $\varepsilon = 0$.
The red, green, black, magenta, and blue lines in all four panels indicate the transitions $|0\rangle \to |1\rangle$,
$|0\rangle \to |2\rangle$, $|0\rangle \to |3\rangle$, $|1\rangle \to |2\rangle$, and $|1\rangle \to |3\rangle$, respectively.
Solid (dashed) lines in panels \textbf{c} and \textbf{d} indicate that the corresponding matrix elements $T_{ij}$ are nonzero (zero).
Allowed and forbidden transitions cross at
$g/2\pi \simeq \omega_{\rm o}/4\pi = 3.15$~GHz\cite{Ashhab10PRA}, where there is an energy-level crossing and the energy eigenstates $|2\rangle$ and $|3\rangle$ exchange their physical states.
The black dotted line is at the coupling strength in circuit~I at $n_{\phi \rm q} = -1.5$,
$g/2\pi=4.92$~GHz.}
\label{wg}
\end{figure}

%accept Htotal
To conclude this analysis of the observed transmission spectra,
the fact that the frequencies of the spectral lines and the points where they become forbidden follow, respectively,
$\omega_{ij}^{\rm cal}$ and $|T_{ij}|$ lends strong support to the conclusion that $\mathcal{H}_{\rm total}$
accurately describes our circuits.
%deep strong coupling
Importantly, in circuits~II and III, $g$ is larger than both $\omega_{\rm o}$ and $\Delta$,
emphasizing that the circuits are in the deep strong-coupling regime [$g \gtrsim \max (\omega_{\rm o},\sqrt{\Delta\omega_{\rm o}}/2)$]\cite{Ashhab10PRA}.
The fact that at $\varepsilon=0$ the two forbidden transitions are located between the two allowed transitions is a further sign that $g>\omega_{\rm o}/2$ (see Fig.~\ref{wg}\textbf{c}).
In contrast, the highest coupling strengths achieved in previous experiments\cite{Niemczyk10, FornDiaz10} give $g/\omega_{\rm o} = 0.12$ and $0.1$, respectively.
%Lamb
From the spectrum in Fig.~3\textbf{a}, we find that
$\omega_{01}(\varepsilon = 0)/\Delta =0.13$~GHz/0.43~GHz$ = 0.30$,
meaning that the Lamb shift\cite{Lamb47PR} is 70\% of the bare qubit frequency.
The same value (0.30) is obtained from theoretical calculations for $g/\omega_{\rm o} = 0.78$.

%no-go
Using our experimental results, we can make a statement regarding the $A^2$ term and the superradiance no-go theorem~\cite{Rzazewski75PRL} in our setup.
A direct consequence of the no-go theorem is that,
provided that the condition of the theorem $(C_{\rm A2}>g/\Delta)$ is satisfied,
the system parameters will be renormalized such that the experimentally measured parameters will satisfy the inequality $2g/\sqrt{\Delta \omega_{\rm o}}<1$ (see Methods).
However, in all five cases in our experiment,
we find that $2g/\sqrt{\Delta \omega_{\rm o}}>1$,
with the ratio on the left-hand side ranging from 2.4 to 9.6 (see Table~I).
These results demonstrate that the $A^2$ term in our setup
does not satisfy the condition of the no-go theorem and therefore does not preclude a superradiant state.
In fact, we expect that $C_{\rm A2}\ll 1$ as shown in Methods. 

\begin{table}
\begin{threeparttable}
\caption{\textbf{The energy eigenstates of the qubit-oscillator system.}}

\begin{tabular}{c@{\hspace{0.5cm}}c@{\hspace{1cm}}c@{\hspace{0.5cm}}c}
    \hline
    \hline
	\multicolumn{2}{l}{energy eigenbasis} & \multicolumn{2}{c}{\hspace{3.0cm}$|\textrm{qubit}\rangle \otimes |\textrm{oscillator}\rangle$ basis}\\
    \hline
$g<\frac{\omega_{\rm o}}{2}$ & $g>\frac{\omega_{\rm o}}{2}$ &arbitrary $g$&$g=0$\\
	\hline
    $|0\rangle$ & $|0\rangle$ & ($|\textrm{\textrm{L}}\rangle_{\rm q}\otimes |-\alpha\rangle_{\rm o} + |\textrm{R}\rangle_{\rm q}\otimes |\alpha\rangle_{\rm o})/\sqrt{2}$ & $|\rm g\rangle_{\rm q}\otimes |0\rangle_{\rm o}$\\
    $|1\rangle$ & $|1\rangle$ & ($|\textrm{\textrm{L}}\rangle_{\rm q}\otimes|-\alpha\rangle_{\rm o} - |\textrm{R}\rangle_{\rm q}\otimes|\alpha\rangle_{\rm o})/\sqrt{2}$ & $|\rm e\rangle_{\rm q}\otimes |0\rangle_{\rm o}$\\
    $|2\rangle$ & $|3\rangle$ & ($|\textrm{\textrm{L}}\rangle_{\rm q}\otimes \hat{D}(-\alpha)|1\rangle_{\rm o} + |\textrm{R}\rangle_{\rm q}\otimes \hat{D}(\alpha)|1\rangle_{\rm o})/\sqrt{2}$ & $|\rm g\rangle_{\rm q}\otimes |1\rangle_{\rm o}$\\
    $|3\rangle$ & $|2\rangle$ & ($|\textrm{\textrm{L}}\rangle_{\rm q}\otimes\hat{D}(-\alpha)|1\rangle_{\rm o} - |\textrm{R}\rangle_{\rm q}\otimes\hat{D}(\alpha)|1\rangle_{\rm o})/\sqrt{2}$ & $|\rm e\rangle_{\rm q}\otimes |1\rangle_{\rm o}$\\
    \hline
    \hline
\end{tabular}
\label{ent}
{\small
\begin{tablenotes}
\item[]The left two columns are written in the energy eigenbasis while the right two columns are written in the tensor product basis of qubit and oscillator states.
At $g \simeq \omega_{\rm o}/2$,
there is an energy-level crossing and the energy eigenstates $|2\rangle$ and $|3\rangle$ exchange their physical states.
$|\textrm{L}\rangle_{\rm q}$ and $|\textrm{R}\rangle_{\rm q}$ are the persistent-current states of the qubit,
$|\textrm{g}\rangle_{\rm q}$ and $|\textrm{e}\rangle_{\rm q}$ are the energy eigenstates of the qubit,
$|\pm\alpha\rangle_{\rm o}=\hat{D}(\pm \alpha)|0\rangle_{\rm o}$ are coherent states of the oscillator,
$\hat{D}(\alpha)$ is a displacement operator,
and $|n\rangle_{\rm o}$ is a Fock state of the bare oscillator.
At $g = 0$ and hence $\alpha = 0$, the energy eigenstates are product states, as shown in the right-most column.
For arbitrary $g$,
the energy eigenstates of the qubit-oscillator system are entangled states.
\end{tablenotes}
}
\end{threeparttable}
\end{table}

%displacement
The energy eigenstates of the qubit-oscillator system can be understood 
in the following way.
In the absence of coupling, the energy eigenstates are product states where the oscillator is described by a Fock state $|n\rangle_{\rm o}$ with $n$ plasmons.
Because of the coupling to the qubit, the
state of the oscillator is displaced in one of two opposite directions depending on the persistent-current state of the qubit\cite{Ashhab10PRA}: $|{\rm L}\rangle_{\rm q}\otimes|n\rangle_{\rm o}\to |{\rm L}\rangle_{\rm q}\otimes\hat{D}(-\alpha)|n\rangle_{\rm o}$ and $|{\rm R}\rangle_{\rm q}\otimes|n\rangle_{\rm o}\to |{\rm R}\rangle_{\rm q}\otimes\hat{D}(\alpha)|n\rangle_{\rm o}$.
Here, $\hat{D}(\alpha)=\exp(\alpha \hat{a}^{\dagger}-\alpha^* \hat{a})$ is the displacement operator,
and $\alpha$ is the displacement.
The amount of the displacement is approximately $\pm g/\omega_{\rm o}$.
As the energy eigenstates of an isolated qubit at $\varepsilon = 0$ are superpositions of the persistent-current states, $|\rm g\rangle_{\rm q} = (|\textrm{L}\rangle_{\rm q} + |\textrm{R}\rangle_{\rm q})/\sqrt{2}$ and $|\rm e\rangle_{\rm q} = (|\textrm{L}\rangle_{\rm q} - |\textrm{R}\rangle_{\rm q})/\sqrt{2}$,
the energy eigenstates of the qubit-oscillator system at $\varepsilon = 0$ are well described by
Schr\"{o}dinger-cat-like entangled states between
persistent-current states of the qubit
and displaced Fock states of the oscillator $\hat{D}(\pm\alpha)|n\rangle_{\rm o}$, as shown in Table~\ref{ent}.
Note that the displaced vacuum state
$\hat{D}(\alpha)|0\rangle_{\rm o}$ is the coherent state
$|\alpha\rangle_{\rm o} = \exp (-|\alpha|^2/2)\sum_{n = 0}^{\infty}\alpha^n/\sqrt{n!}|n\rangle_{\rm o}$.
Although the above picture works best when $\omega_{\rm o}\gg\Delta$,
theoretical calculations show that it also gives a rather accurate description for circuit~III
(with $\omega_{\rm o}/\Delta = 1.44$) (see Methods).
The vanishing of the spectral lines corresponding to the $\ket{0}\rightarrow\ket{2}$ and $\ket{1}\rightarrow\ket{3}$ transitions at $\varepsilon=0$ is a consequence of the symmetric form of the
energy eigenstates.
This symmetry is expected from the current-inversion symmetry in the Hamiltonian $\mathcal{H}_{\rm total}$,
and it supports the theoretical prediction that the energy eigenstates at that point are qubit-oscillator entangled states.

%numerically calculate gs + te entanglement
Using $\mathcal{H}_{\rm total}$ and the parameters shown in Table~\ref{params},
we can calculate the qubit-oscillator ground-state entanglement
$\mathcal{E}_{\rm gs}$ (see Supplementary Information, section~S5).
In all cases, $\mathcal{E}_{\rm gs}\gtrsim 90\%$, and for circuit II in particular $\mathcal{E}_{\rm gs} = 99.88\%$.
In comparison, the ground-state entanglement for the parameters of Refs. 12 and 13 is 6\% and 4\%, respectively.
It should be noted here that in all five cases in our experiment there will be a significant population in the state $|1\rangle$ in thermal equilibrium,
and the thermal-equilibrium qubit-oscillator entanglement will be reduced to below 8\% for circuits I and II, and 25\% for circuit III (see Supplementary Information, Table~S1).

%conclusion
In conclusion, we have experimentally achieved deep-strong coupling between a superconducting flux qubit and an LC oscillator.
Our results are consistent with the theoretical prediction that the energy eigenstates are Schr\"{o}dinger-cat-like entangled states between persistent-current states of the qubit and displaced Fock states of the oscillator.
We have also observed a huge Lamb shift, 70\% of the bare qubit frequency.
The tiny Lamb shift in natural atoms, which arises from weak vacuum fluctuations, was one of the earliest phenomena to stimulate the study of quantum electrodynamics. Now we can design artificial systems with light-matter interaction so strong that instead of speaking of vacuum fluctuations we speak of a strongly correlated light-matter ground state, defining a new state of matter and opening prospects for applications in quantum technologies.

%note added in proof
Note added in proof: After acceptance of our paper, we became aware of a related manuscript (Ref.~27)
taking a different approach to the same theme.

\begin{acknowledgments}
We thank Kae Nemoto, Masao Hirokawa, Kunihiro Inomata, John W. Munro, Yuichiro Matsuzaki, 
Motoaki Bamba, and Norikazu Mizuochi for stimulating discussions.
The authors are greatful to Mikio Fujiwara, Kentaro Wakui, Masahiro Takeoka, and Masahide Sasaki for their continued support through all the stages of this research.
We thank Junichi Komuro, Shinya Inoue, and Etsuro Sasaki for assistance with experimental setup. 

We also thank Sander Weinreb for their support by providing excellent cryoamplifiers,
and Noriyoshi Matsuura and Yoshitada Kato for their cordial support in the startup phase of this research.
Some of our calculations were performed using the QuTiP simulation package~\cite{QuTiP}.
 
This work was supported in part by the Scientific Research (S) Grant No. JP25220601 by the Japanese Society for the Promotion of Science (JSPS). 
\end{acknowledgments}

\section*{Author contributions}
All authors contributed extensively to the work presented in this paper. 
F. Y., T. F., K. S. carried out measurements and data analysis on the coupled flux qubit - LC-oscillator circuits.
F. Y., T. F. designed and F. Y., T. F., K. K. fabricated the flux-qubit and associated devices.
T. F., F. Y., K. K., S. S., and K. S. designed and developed the measurement system.
%W. J. M., A. K., Y. T., H. N., M. S. E. and K. N. provided theoretical support and analysis.
S. A. provided theoretical support and analysis.
F. Y., T. F., S. A., and K. S. wrote the manuscript, with feedback from all authors.
K. S. designed and supervised the project.

\section*{METHODS}
\noindent
\textbf{Laser microscope image.}
The laser microscope image in Fig.~1\textbf{b} was obtained by Keyence VK-9710 Color 3D Laser Scanning Microscope.
The magnification of the objective lens is 10.
The application ``VK Viewer'' was used for image acquisition.

\noindent
\textbf{Scanning electron microscope image.}
The scanning electron microscope image in Fig.~1\textbf{c} was obtained by JEOL JIB-4601F.
The acceleration voltage was 10~kV,
the magnification was 6500,
and the working distance was 8.7 mm.

\noindent
\textbf{Nonlinearity of $M$ and the $A^2$ term of the total Hamiltonian.}
We now consider the nonlinearity of the mutual inductance $M$ between the flux qubit and the LC oscillator.
As discussed in the Supplementary Information,
$M$ is almost the same as $L_{\rm c}$ in Fig.~1\textbf{a}, which depends on the current flowing through the Josephson junction $I_{\rm b}$ as $L_{\rm c}(I_{\rm b})=\Phi_0/(2\pi \sqrt{(a_{\rm c}I_{\rm c})^2-I_{\rm b}^2})$,
where $a_{\rm c}I_{\rm c}\equiv I_{\rm cM}$ is the critical current of the Josephson junction.
We thus assume that $M$ can similarly be written as

\begin{eqnarray}
\label{MIb}
M(I_{\rm b})=\frac{\Phi_0}{2\pi \sqrt{I_{\rm cM}^2-I_{\rm b}^2}}.
\end{eqnarray}
The nonlinearity of $M(I_{\rm b})$ up to second order in $\delta I_{\rm b}$ can be written as

\begin{align}
\nonumber
M(I_{\rm b} + \delta I_{\rm b})&=M(I_{\rm b}) + \delta I_{\rm b}\frac{\partial M(I_{\rm b})}{\partial I_{\rm b}} + \frac{\delta I_{\rm b}^2}{2}\frac{\partial^2 M(I_{\rm b})}{\partial I_{\rm b}^2}\\
&=M(I_{\rm b})\left ( 1 + \frac{I_{\rm b}\delta I_{\rm b}}{I_{\rm cM}^2 - I_{\rm b}^2} + \frac{I_{\rm cM}^2 + 2I_{\rm b}^2}{2(I_{\rm cM}^2-I_{\rm b}^2)^2}\delta I_{\rm b}^2\right ).
\end{align}

The coupling Hamiltonian can be written as $\mathcal{H}_{\rm c} = M(\hat{I}_{\rm q} + \hat{I}_{\rm o})\hat{I}_{\rm q}\hat{I}_{\rm o} = M(\hat{I}_{\rm q} + \hat{I}_{\rm o})I_{\rm p}\sigma_z I_{\rm zpf}(\hat{a}+\hat{a}^{\dagger})$,
where $\hat{I}_{\rm q} = I_{\rm p}\sigma_z$ is the persistent-current operator of the qubit,
$\hat{I}_{\rm o} = I_{\rm zpf}(\hat{a}+\hat{a}^{\dagger})$ is the current operator of the oscillator,
and the current $\hat{I}_{\rm q} + \hat{I}_{\rm o}$ flows through the mutual inductance.
Typically, $I_{\rm p}\gg I_{\rm zpf}$.
Taking into account the nonlinearity of $M(\hat{I}_{\rm q} + \hat{I}_{\rm o})$,
the coupling Hamiltonian is written as

\begin{align}
\nonumber
\mathcal{H}_{\rm c} &= M(\hat{I}_{\rm q} + \hat{I}_{\rm o})\hat{I}_{\rm q}\hat{I}_{\rm o}\\
\nonumber
&=M(\hat{I}_{\rm q})\left ( 1 + \frac{\hat{I}_{\rm q}\hat{I}_{\rm o}}{I_{\rm cM}^2 - \hat{I}_{\rm q}^2} + \frac{I_{\rm cM}^2 + 2\hat{I}_{\rm q}^2}{2(I_{\rm cM}^2-\hat{I}_{\rm q}^2)^2}\hat{I}_{\rm o}^2\right )\hat{I}_{\rm q}\hat{I}_{\rm o}\\
\nonumber
&=M(I_{\rm p})\left [I_{\rm p}I_{\rm zpf}\sigma_z(\hat{a} + \hat{a}^{\dagger}) + \frac{I_{\rm p}^2I_{\rm zpf}^2}{I_{\rm cM}^2-I_{\rm p}^2}(\hat{a} + \hat{a}^{\dagger})^2 + \frac{(I_{\rm cM}^2 + 2I_{\rm p}^2)I_{\rm p}I_{\rm zpf}^3}{2(I_{\rm cM}^2-I_{\rm p}^2)^2}\sigma_z(\hat{a} + \hat{a}^{\dagger})^3\right ]\\
\label{Hc2nd}
&=\hbar g[\sigma_z(\hat{a} + \hat{a}^{\dagger}) + C_{\rm A2}(\hat{a} + \hat{a}^{\dagger})^2 + C_{\rm A3}\sigma_z(\hat{a} + \hat{a}^{\dagger})^3],
\end{align}
where

\begin{eqnarray}
\label{gMII}
\hbar g = M(I_{\rm p})I_{\rm p}I_{\rm zpf},
\end{eqnarray}

\begin{eqnarray}
\label{Ca2}
C_{\rm A2} = \frac{I_{\rm p}I_{\rm zpf}}{I_{\rm cM}^2-I_{\rm p}^2},
\end{eqnarray}
and

\begin{eqnarray}
\label{Ca3}
C_{\rm A3} = \frac{(I_{\rm cM}^2 + 2I_{\rm p}^2)I_{\rm zpf}^2}{2(I_{\rm cM}^2-I_{\rm p}^2)^2}.
\end{eqnarray}
Here, we considerd terms up to second order in $I_{\rm zpf}/I_{\rm p}$.
We find that  $1 \gg C_{\rm A2} \gg C_{\rm A3}$ considering the following relation,
$I_{\rm cM}(= a_{\rm c}I_{\rm c}) > I_{\rm p}(\lesssim a_3I_{\rm c}) \gg I_{\rm zpf}(\ll I_{\rm c})$,
where $a_{\rm c} \gtrsim 1$ (see Supplementary Material),
$0.4 \lesssim a_3 \lesssim 0.8$,
$I_{\rm c}$ is several hundred nano amperes,
and $I_{\rm zpf}$ is several ten nano amperes.
Since the term $C_{\rm A3}$ is very small, we ignore the third term in Eq.~(\ref{Hc2nd}).

%total Hamiltonian with A^2 term
The total Hamiltonian of the circuit considering the nonlinearity of $M$ up to first order in $I_{\rm zpf}/I_{\rm p}$ is given by

\begin{eqnarray}
\label{HtotalA2}
\mathcal{H}_{\rm total} &= -\frac{\hbar}{2}(\Delta \sigma_x + \varepsilon \sigma_z) + \hbar \omega_{\rm o}\left ( \hat{a}^{\dagger}\hat{a} + \frac{1}{2}\right ) + \hbar g\sigma_z(\hat{a} + \hat{a}^{\dagger}) + C_{\rm A2}\hbar g(\hat{a} + \hat{a}^{\dagger})^2,
\end{eqnarray}
where the first term is the Hamiltonian of the flux qubit, the second term is the Hamiltonian of the LC oscillator, and the third term is the coupling Hamiltonian.
The fourth term proportional to $(\hat{a} + \hat{a}^{\dagger})^2$ is known as the $A^2$ term in atomic physics.
This term can be eliminated by a variable transformation as

\begin{align}
\nonumber
\mathcal{H}_{\rm total} &= -\frac{\hbar}{2}(\Delta \sigma_x + \varepsilon \sigma_z) + \hbar \omega_{\rm o}\left ( \hat{a}^{\dagger}\hat{a}+\frac{1}{2}\right ) + C_{\rm A2}\hbar g(\hat{a} + \hat{a}^{\dagger})^2 + \hbar g\sigma_z(\hat{a} + \hat{a}^{\dagger})\\
\nonumber
&= - \frac{\hbar}{2}(\Delta \sigma_x + \varepsilon \sigma_z) + \left ( \frac{\hbar \omega_{\rm o}}{4} + C_{\rm A2}\hbar g\right )(\hat{a} + \hat{a}^{\dagger})^2 - \frac{\hbar \omega_{\rm o}}{4}(\hat{a} - \hat{a}^{\dagger})^2 + \hbar g\sigma_z(\hat{a} + \hat{a}^{\dagger})\\
\nonumber
&= - \frac{\hbar}{2}(\Delta \sigma_x + \varepsilon \sigma_z) + \frac{\hbar \omega_{\rm o}'}{4}(\hat{b} + \hat{b}^{\dagger})^2 - \frac{\hbar \omega_{\rm o}'}{4}(\hat{b} - \hat{b}^{\dagger})^2 + \hbar g'\sigma_z(\hat{b} + \hat{b}^{\dagger})\\
\label{Htotalb}
&= - \frac{\hbar}{2}(\Delta \sigma_x + \varepsilon \sigma_z) + \hbar \omega_{\rm o}'(\hat{b}^{\dagger}\hat{b} + \frac{1}{2}) + \hbar g'\sigma_z(\hat{b} + \hat{b}^{\dagger}),
\end{align}
where

\begin{eqnarray}
\label{omegao}
\omega_{\rm o}' = \sqrt{\omega_{\rm o}^2+4C_{\rm A2}g\omega_{\rm o}},
\end{eqnarray}

\begin{eqnarray}
\label{go}
g' = \sqrt{\frac{\omega_{\rm o}}{\omega_{\rm o}'}} g,
\end{eqnarray}
and the new field operators,
\begin{eqnarray}
\label{bpb}
\hat{b} + \hat{b}^{\dagger} = \sqrt{\frac{\omega_{\rm o}'}{\omega_{\rm o}}}(\hat{a} + \hat{a}^{\dagger})
\end{eqnarray}
and
\begin{eqnarray}
\label{bmb}
\hat{b} - \hat{b}^{\dagger} = \sqrt{\frac{\omega_{\rm o}}{\omega_{\rm o}'}}(\hat{a} - \hat{a}^{\dagger}),
\end{eqnarray}
are used.
%Note that with this transformation, $[\hat{b},\hat{b}^{\dagger}] = [\hat{a},\hat{a}^{\dagger}] = 1$.
%Commutation relations for the $\hat{b}$ operator remain the same as those for the $\hat{a}$ operator.
The form of the Hamiltonian in Eq.~(\ref{Htotalb}) is exactly the same as the one where the coupling term is linear in $(\hat{a} + \hat{a}^{\dagger})$, which is given by
\begin{eqnarray}
\label{Htotal_linear}
\mathcal{H}_{\rm total}^{\rm linear} &= -\frac{\hbar}{2}(\Delta \sigma_x + \varepsilon \sigma_z) + \hbar \omega_{\rm o}(\hat{a}^{\dagger}\hat{a} + \frac{1}{2}) + \hbar g\sigma_z(\hat{a} + \hat{a}^{\dagger}).
\end{eqnarray}

Note that the transformation described by Eqs.~(\ref{bpb}) and (\ref{bmb}) is a Hopfield-Bogoliubov transformation\cite{{Hopfield58PR}}.
It guarantees that $[\hat{b},\hat{b}^{\dagger}]=[\hat{a},\hat{a}^{\dagger}]=1$.
In other words, both the $\hat{a}$ operators and the $\hat{b}$ operators obey the harmonic oscillator commutation relations.
The two sets of operators are related to each other by quadrature squeezing operations.
The most natural choice among these two and all other quadrature-squeezed variants is the one that leads to the standard form of the harmonic oscillator Hamiltonian, usually expressed as $\hbar \omega_{\rm o} \hat{a}^{\dagger} \hat{a}$.
As such, the $\hat{b}$ operators are the most natural oscillator operators for our circuits.
The $\hat{a}$ operators were defined based on an incomplete description of the circuit, considering the properties of the LC circuit and ignoring the qubit and coupler parts of the circuit.
In particular, the $A^2$ term in our circuits describes an additional contribution to the inductive energy of the oscillator that arises in the presence of the qubit and coupler circuits.
Similarly, the expression given in the main text for the current zero-point fluctuations must be modified in order to correctly describe the fluctuations in the full circuit.

\noindent
\textbf{Condition for superradiant phase transition.}
In cases where one expects a sharp transition from a normal to a superradiant state, e.g.~when $\Delta\gg\omega_{\rm o}$ or when the single qubit is replaced by a large ensemble of $N$ qubits (and $g$ is defined to include the ensemble enhancement factor $\sqrt{N}$), the phase transition condition (without the $A^2$ term) is:
\begin{equation}
4 g^2 = \Delta \times \omega_{\rm o}.
\end{equation}
After taking into account the renormalization of $\omega_{\rm o}$ and $g$ caused by the $A^2$ term as described above, the condition for the phase transition becomes
\begin{equation}
4 g^2 \sqrt{\frac{\omega_{\rm o}}{\omega_{\rm o}+4C_{\rm A2}g}} = \Delta \times \omega_{\rm o} \times \sqrt{\frac{\omega_{\rm o}+4C_{\rm A2}g}{\omega_{\rm o}}},
\end{equation}
or in other words
\begin{equation}
4 g^2 = \Delta \times \left(\omega_{\rm o}+4C_{\rm A2}g\right).
\end{equation}
If
%, for some reason, $\kappa$ is related to $g$ (e.g.~because they result from the same microscopic mechanism) and
the parameters are constrained to satisfy the relation $C_{\rm A2}>g/\Delta$, the right-hand side increases whenever we increase the left-hand side, and no matter how large $g$ becomes it will never be strong enough to satisfy the phase transition condition.
This can indeed be the case with atomic qubits, and it leads to the no-go theorem in those systems$^{16}$.%\cite{Rzazewski75PRL}.
%\subsection*{parameter estimation}
% M,Ip,Izpf

\noindent
\textbf{Fidelities of qubit-oscillator entangled states for circuit~III.}
The fidelity between two pure states $|\phi\rangle$ and $|\psi\rangle$ is given by
$F(|\phi\rangle,\,|\psi\rangle) = |\langle \phi | \psi \rangle|^2$.
For circuit~III, the fidelities between the four lowest energy eigenstates given in Table~II $|i_{\rm TII}\rangle$ and the corresponding exact energy eigenstates of $\mathcal{H}_{\rm total}$ $|i_{\rm exact}\rangle$ ($i = 0$, 1, 2, 3) are calculated to be
$F(|0_{\rm TII}\rangle,\,|0_{\rm exact}\rangle)$ = 0.981,
$F(|1_{\rm TII}\rangle,\,|1_{\rm exact}\rangle)$ = 0.985,
$F(|2_{\rm TII}\rangle,\,|2_{\rm exact}\rangle)$ = 0.975,
and $F(|3_{\rm TII}\rangle,\,|3_{\rm exact}\rangle)$ = 0.967.
All the other data sets give significantly higher fidelities. In
particular, for circuit II $F(|0_{\rm TII}\rangle,\,|0_{\rm exact}\rangle)=0.99994$.

\section*{Supplementary Information}
\setcounter{figure}{0}
\renewcommand\theequation{S\arabic{equation}}
\renewcommand\thesection{S\arabic{section}}
\renewcommand\thefigure{S\arabic{figure}}
\renewcommand\thetable{S\Roman{table}}
\section{Flux bias dependence of the coupler's critical current}
%We start from the Josephson junction circuit of the flux qubit including the coupler shown in Fig.~\ref{FQCirc}\textbf{a}, which contains all the components of circuit~I,
%and then show that the circuit can be simplified to the ones in Fig.~\ref{FQCirc}\textbf{b} and \textbf{c}.
\begin{figure}
\includegraphics{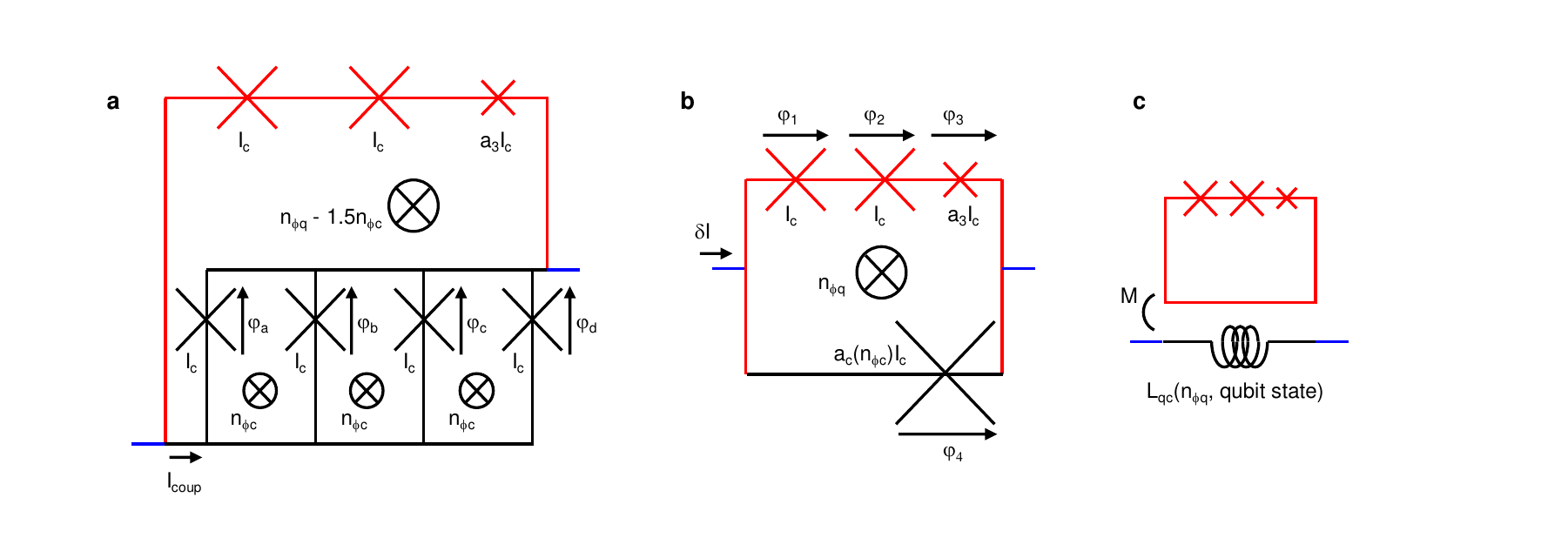}
\caption{\textbf{Circuit diagrams of the flux qubit and coupler.}
\textbf{a},~The qubit (red and black) consists of three Josephson junctions in the upper branch (red) and the coupler (black), which is four parallel Josephson junctions.
\textbf{b},~The coupler junctions are simplified to a single effective Josephson junction.
\textbf{c},~The equivalent circuit of \textbf{b}, now consisting of the mutual inductance $M$ and the inductance across the qubit and the coupler $L_{\rm qc}$, which depends on both the flux bias and the qubit state.
\label{FQCirc}
}
\end{figure}

The circuit diagram of the coupler in circuit~I is shown as the black part of the circuit in Fig.~\ref{FQCirc}\textbf{a}.
Here, $n_{\phi \rm c}$ is the normalized flux bias in units of the superconducting flux quantum $\Phi_0 = h/2e$ through each coupler loop defined by two neighboring parallel junctions.
The critical currents of the two large junctions of the flux qubit and the four junctions of the coupler are all approximately equal, with the value $I_{\rm c}$.
The current through the coupler $I_{\rm coup}$ is the sum of the currents across the four Josephson junctions: $I_{\rm coup} = I_{\rm c}(\sin\varphi_{\rm a} + \sin\varphi_{\rm b} + \sin\varphi_{\rm c} + \sin\varphi_{\rm d})$, where $\varphi_i$ ($i$ = a, b, c, d) is the phase across junction $i$. 
Considering the fluxoid quantization of each loop,
the phases can be written using $\varphi_{\rm a}$ and $n_{\phi \rm c}$ as

\begin{eqnarray}
\label{phib}
\varphi_{\rm b}=\varphi_{\rm a} + 2\pi n_{\phi \rm c},\\
\varphi_{\rm c}=\varphi_{\rm a} + 4\pi n_{\phi \rm c},
\end{eqnarray}
and

\begin{eqnarray}
\label{phid}
\varphi_{\rm d}=\varphi_{\rm a} + 6\pi n_{\phi \rm c}.
\end{eqnarray}
Here, we ignore the sum of the kinetic and geometric inductances of the superconducting lead, which is at least an order of magnitude smaller than those of the Josephson junctions.
Using Eqs.~(\ref{phib})--(\ref{phid}), $I_{\rm coup}$ can be written as

\begin{align}
\label{phia}
\nonumber
I_{\rm coup}& = I_{\rm c}[\sin\varphi_{\rm a} + \sin (\varphi_{\rm a} + 2\pi n_{\phi \rm c}) + \sin(\varphi_{\rm a} + 4\pi n_{\phi \rm c}) + \sin(\varphi_{\rm a} + 6\pi n_{\phi \rm c})]\\
\nonumber
& = 2I_{\rm c}[\sin(\varphi_{\rm a} + \pi n_{\phi \rm c})\cos(\pi n_{\phi \rm c}) + \sin(\varphi_{\rm a} + 5\pi n_{\phi \rm c})\cos(\pi n_{\phi \rm c})]\\
&= 4I_{\rm c}\sin(\varphi_{\rm a} + 3\pi n_{\phi \rm c}) \cos (2\pi n_{\phi \rm c})\cos (\pi n_{\phi \rm c}).
\end{align}
Thus, the critical current of the coupler $I_{\rm c(coup)}$ can be described by the ratio
\begin{eqnarray}
\label{IccI}
a_{\rm c}(n_{\phi \rm c}) = \frac{I_{\rm c(coup)}}{I_{\rm c}} = 4\cos (2\pi n_{\phi \rm c})\cos (\pi n_{\phi \rm c}).
\end{eqnarray}
Now, the coupler junctions in Fig.~\ref{FQCirc}\textbf{a} can be replaced by a single effective Josephson junction whose critical current is $a_{\rm c}(n_{\phi \rm c})I_{\rm c}$ as shown in Fig.~\ref{FQCirc}\textbf{b}.
Circuit~II is almost the same as circuit~I except that its coupler consists of two Josephson junctions of critical current $I_{\rm c}$,
forming a superconducting quantum interference device (SQUID). 
The critical-current ratio of the SQUID is described by
\begin{eqnarray}
\label{acsq}
a_{\rm cII}(n_{\phi \rm cII})=2\cos (\pi n_{\phi \rm cII}),
\end{eqnarray}
where $n_{\phi {\rm cII}}$ is the normalized flux bias through the SQUID loop.
Thus, the  circuit diagram of the flux qubit in circuit~II is also described by Fig.~\ref{FQCirc}\textbf{b}.

\section{Estimation of $L_{\rm qc}$ and $M$}
The circuit in Fig.~\ref{FQCirc}\textbf{b} should be simplified to the one in Fig.~\ref{FQCirc}\textbf{c} to estimate $L_{\rm qc}$ and $M$ as functions of the bias current $\delta I$
coming from the current in the LC oscillator and normalized flux bias through the qubit loop $n_{\phi \rm q}$ in units of $\Phi_0$.
The total Josephson energy of the circuit is given by
\begin{eqnarray}
E_{\rm J}^{\rm total} = -E_{\rm J}\left [ \cos\varphi_1 + \cos\varphi_2 + a_3\cos\varphi_3 + a_{\rm c}\cos(-\varphi_{\rm u}+2\pi n_{\phi \rm q})\right ] -\frac{\delta I\Phi_0}{2\pi}\varphi_{\rm x},
\end{eqnarray}
where $E_{\rm J} = \Phi_0 I_{\rm c}/2\pi$, $\varphi_i$ ($i$ = 1, 2, 3) is the phase difference across the $i$th junction,
$a_3$ and $a_{\rm c}$ are the critical current ratios of the third and the coupler junctions,
$\varphi_{\rm u} = \varphi_1+\varphi_2 + \varphi_3$ is the phase difference across the upper branch of the qubit loop,
and $\varphi_{\rm x} = (\varphi_{\rm u} + \pi n_{\phi \rm q})$ is the average phase difference across the upper and lower branches of the qubit loop.
The last term is the energy of the bias current source.

\begin{figure}
\includegraphics{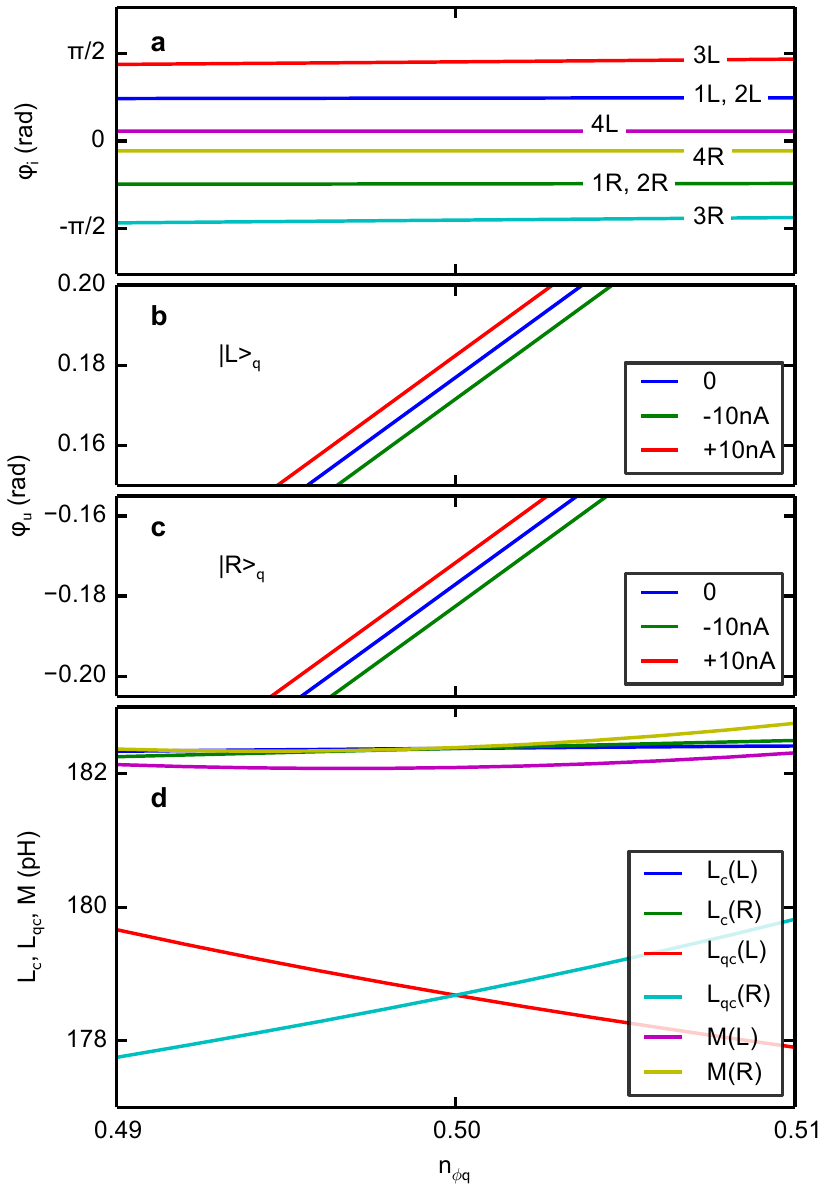}
\caption{\textbf{Flux bias dependence of phases and inductances.}
\textbf{a},~The flux bias dependence of the phase across the different Josephson junctions in Fig.~\ref{FQCirc}\textbf{b} when the qubit state is $|\rm L\rangle_{\rm q}$ and $|\rm R\rangle_{\rm q}$.
\textbf{b} (\textbf{c}), The flux bias dependence of the phase across the upper branch of the qubit loop $\varphi_{\rm u}$ at three different current bias values, $\delta I=0,\,\pm10$~nA, when the qubit state is $|\rm L\rangle_{\rm q}$ ($|\rm R\rangle_{\rm q}$).
\textbf{d},~The flux bias dependence of the coupler inductance $L_{\rm c}$, the mutual inductance $M$, and the inductance across the qubit and the coupler $L_{\rm qc}$ when the qubit state is $|\rm L\rangle_{\rm q}$ and $|\rm R\rangle_{\rm q}$.
\label{LMn}
}
\end{figure}

% phase at each local minimum
At $n_{\phi \rm q}\sim 0.5$, $E_{\rm J}^{\rm total}$ has two local minima in the three-dimensional parameter space spanned by $\varphi_1$, $\varphi_2$, and $\varphi_3$.
The localized state at each minimum corresponds to one of the two persistent-current states of the flux qubit, $|\rm L\rangle_{\rm q}$ and $|\rm R\rangle_{\rm q}$.
For simplicity, we use the sets of phases $\{\varphi_{i}^{|\rm L\rangle} \}$ and $\{\varphi_{i}^{|\rm R\rangle} \}$ at the minima of $E_{\rm J}^{\rm total}$ as the values of the different phases for $|\rm L\rangle_{\rm q}$ and $|\rm R\rangle_{\rm q}$.
Figure~\ref{LMn}\textbf{a} shows the $n_{\phi \rm q}$ dependence of different phases corresponding to $|\rm L\rangle_{\rm q}$ and $|\rm R\rangle_{\rm q}$.
In the calculation, the parameters $I_{\rm c} = 460$~nA and $a_3 = 0.705$, which are estimated from other samples fabricated simultaneously in the same fabrication process, are used.
We also assume that the global magnetic field simultaneously provides flux bias through the loops of the qubit and the coupler according to their area ratio as $n_{\phi \rm q}:n_{\phi \rm c} =24:1$.

The qubit-state-dependent inductance across the qubit and coupler $L_{\rm qc}^{|\rm L(R)\rangle}$ is calculated considering the Josephson inductances, $L_{\textrm{J}1}^{|\rm L(R)\rangle}=\Phi_0/(2\pi I_{\rm c}\cos \varphi_1^{|\rm L(R)\rangle})$, $L_{\textrm{J}2}^{|\rm L(R)\rangle}=\Phi_0/(2\pi I_{\rm c}\cos \varphi_2^{|\rm L(R)\rangle})$, $L_{\textrm{J}3}^{|\rm L(R)\rangle}=\Phi_0/(2\pi a_3 I_{\rm c}\cos \varphi_3^{|\rm L(R)\rangle})$, and $L_{\textrm{J}4}^{|\rm L(R)\rangle}=\Phi_0/(2\pi a_{\rm c}I_{\rm c}\cos \varphi_4^{|\rm L(R)\rangle})$, as

\begin{eqnarray}
L_{\rm qc}^{|\rm L(R)\rangle} = \frac{L_{\textrm{J}4}^{|\rm L(R)\rangle}(L_{\textrm{J}1}^{|\rm L(R)\rangle} + L_{\textrm{J}2}^{|\rm L(R)\rangle} + L_{\textrm{J}3}^{|\rm L(R)\rangle})}{L_{\textrm{J}1}^{|\rm L(R)\rangle} + L_{\textrm{J}2}^{|\rm L(R)\rangle} + L_{\textrm{J}3}^{|\rm L(R)\rangle} + L_{\textrm{J}4}^{|\rm L(R)\rangle}}.
\end{eqnarray}
Figure~\ref{LMn}\textbf{d} shows the flux-bias dependence of $L_{\rm qc}^{|\rm L\rangle}$ and $L_{\rm qc}^{|\rm R\rangle}$, which can be approximately described as $L_{\rm qc}^{|\textrm{L}\rangle} = L_{\rm qc0} + D_{\rm L}^{|\textrm{L}\rangle}(n_{\phi \rm q}-0.5)$ and
$L_{\rm qc}^{|\textrm{R}\rangle} = L_{\rm qc0} - D_{\rm L}^{|\textrm{R}\rangle}(n_{\phi \rm q}-0.5)$ ($D_{\rm L}^{|\textrm{L}\rangle} \sim D_{\rm L}^{|\textrm{R}\rangle}<0$),
respectively.
The small asymmetry between $L_{\rm qc}^{|\rm L\rangle}$ and $L_{\rm qc}^{|\rm R\rangle}$ is due to the flux-bias dependence of $a_{\rm c}(n_{\phi c})$.
Note that at $n_{\phi \rm q}=0.5$, $L_{\rm qc}^{|\rm L\rangle} = L_{\rm qc}^{|\rm R\rangle} = L_{\rm qc0}$.
The inductances of the coupler junction, $L_{\rm c}^{|\rm L(R)\rangle} = L_{\rm J4}^{|\rm L(R)\rangle}$ are also plotted in Fig.~\ref{LMn}\textbf{d}, and are slightly larger than $L_{\rm qc}^{|\rm L(R)\rangle}$.

\begin{figure}
\includegraphics{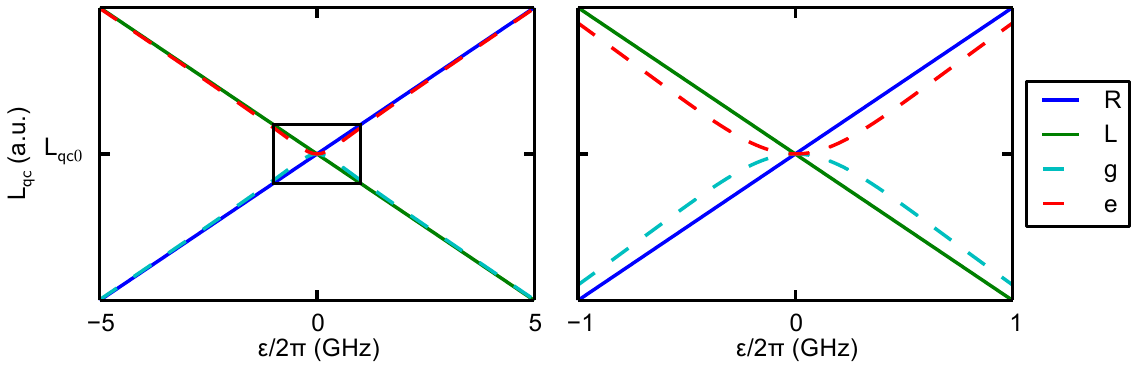}
\caption{\textbf{Flux-bias and qubit-state dependences of $L_{\rm qc}$.}
The right panel is the magnification of the rectangle part of the left panel.
The blue and green solid lines and the red and cyan dashed lines correspond to $L_{\rm qc}^{|\rm R\rangle}$, $L_{\rm qc}^{|\rm L\rangle}$, $L_{\rm qc}^{|\rm g\rangle}$, and $L_{\rm qc}^{|\rm e\rangle}$, respectively.
}
\label{L_qc}
\end{figure}

%Lc,omega_r
It is more convenient to describe the qubit-state-dependent inductance using the energy eigenstates of the qubit, $|\rm g\rangle_{\rm q}$ and $|\rm e\rangle_{\rm q}$, as

\begin{align}
\label{Lqc}
\nonumber
L_{\rm qc} &= \frac{1}{2}(L_{\rm qc}^{|\textrm{g}\rangle}+L_{\rm qc}^{|\textrm{e}\rangle})+\frac{1}{2}(L_{\rm qc}^{|\textrm{g}\rangle}-L_{\rm qc}^{|\textrm{e}\rangle})\sigma_z^{\rm eig}\\
&=\frac{1}{2}(L_{\rm qc}^{|\textrm{g}\rangle}+L_{\rm qc}^{|\textrm{e}\rangle})+\frac{1}{2}(L_{\rm qc}^{|\textrm{g}\rangle}-L_{\rm qc}^{|\textrm{e}\rangle})(\cos\theta\sigma_z + \sin\theta\sigma_x),
\end{align}
where $\sigma_z^{\rm eig}$ is Pauli matrix in the energy eigenbasis,
$\sigma_{x,\,z}$ are Pauli matrices in the persistent-current basis,
$\theta$ is defined as $\cos\theta = \varepsilon/\sqrt{\Delta^2 + \varepsilon^2}$,
and $L_{\rm qc}^{|\textrm{g(e)}\rangle}$ is the inductance across the qubit and the coupler when the qubit state is $|\textrm{g(e)}\rangle_{\rm q}$.
The relation between the persistent-current states and the energy eigenstates of the qubit is written as
\begin{eqnarray}
\label{qgL}
\begin{pmatrix}
|\textrm{g}\rangle_{\rm q} \\
|\textrm{e}\rangle_{\rm q}
\end{pmatrix}
=
\begin{pmatrix}
\cos\frac{\theta}{2} & \sin\frac{\theta}{2} \\
\sin\frac{\theta}{2} & -\cos\frac{\theta}{2}
\end{pmatrix}
\begin{pmatrix}
|\textrm{L}\rangle_{\rm q} \\
|\textrm{R}\rangle_{\rm q}
\end{pmatrix}
.
\end{eqnarray}
Thus $L_{\rm qc}^{|\textrm{L(R)}\rangle}$ can be transformed to $L_{\rm qc}^{|\textrm{g(e)}\rangle}$ as
%and  is calculated using the transformation matrix, where each matrix element is the square of those in Eq.~\ref{qgL}:
\begin{eqnarray}
\label{Lclr}
\begin{pmatrix}
L_{\rm qc}^{|\textrm{g}\rangle} \\
L_{\rm qc}^{|\textrm{e}\rangle}
\end{pmatrix}
=
\begin{pmatrix}
\cos^2\frac{\theta}{2} & \sin^2\frac{\theta}{2} \\
\sin^2\frac{\theta}{2} & \cos^2\frac{\theta}{2}
\end{pmatrix}
\begin{pmatrix}
L_{\rm qc}^{|\textrm{L}\rangle} \\
L_{\rm qc}^{|\textrm{R}\rangle}
\end{pmatrix}
.
\end{eqnarray}
$L_{\rm qc}^{|\rm L(R)\rangle}$ and $L_{\rm qc}^{|\rm g(e)\rangle}$
are shown in Fig.~\ref{L_qc} as functions of the energy bias $\varepsilon$:
$L_{\rm qc}^{|\rm L\rangle}$ and $L_{\rm qc}^{|\rm R\rangle}$ are straight lines,
while $L_{\rm qc}^{|\rm g\rangle}$ and $L_{\rm qc}^{|\rm e\rangle}$ are $\Lambda$-shaped and V-shaped, respectively.
Note that the resonance frequency of the LC oscillator $\omega_{\rm o}=\frac{1}{\sqrt{(L_0 + L_{\rm qc})C}}$ also depends on the qubit state and the flux bias via $L_{\rm qc}$ except at $\varepsilon = 0$.
At sufficiently low temperatures, the qubit is in the ground state,
and $\omega_{\rm o}$ as a function of $\varepsilon$ is V shaped.
%$\omega_{\rm o}(\varepsilon = 0) = \omega_{\rm o0}=\frac{1}{\sqrt{(L_{0}+L_{\rm qc0})C}}$.

% M
The mutual inductance between the qubit loop and the LC oscillator
$M$ can be calculated as $M=\Phi_0|\delta n_{\phi \rm q}/\delta I|$,
where $(\partial \varphi_{\rm u}/\partial n_{\phi \rm q})\delta n_{\phi \rm q} = [\varphi_{\rm u}(\delta I)-\varphi_{\rm u}(-\delta I)]/2$. 
%$\delta n_{\phi \rm q}$ is the shift of $\varphi_{\rm u}$ in the $\varphi - n_{\phi \rm q}$ plane by the small bias current across the qubit and coupler $\delta I$.
The phase $\varphi_{\rm u}$ for $|\rm L\rangle$ ($|\rm R\rangle$) as a function of $n_{\phi \rm q}$ at $\delta I = \pm10$~nA is shifted from that at $\delta I = 0$ as shown in Fig.~\ref{LMn}\textbf{b} (\textbf{c}).
From the shifts, $M$ is obtained as shown in Fig.~\ref{LMn}\textbf{d}.
$M$ is found to be very close to the coupler inductance $L_{\rm c}$.
The flux bias dependence of $a_{\rm c}(n_{\phi \rm c})$ causes a small difference in $M$ between two cases of $|\rm L\rangle_{\rm q}$ and $|\rm R\rangle_{\rm q}$, which is less than 1~$\%$ and we ignore it in the analysis in the main text and the consideration of the nonlinearity of $M$ in Methods.
%\subsection*{Nonlinearity of $M$ and the $A^2$ term of the total Hamiltonian}
%nonlinear M

\section{Measurement setup}
\begin{figure}
\includegraphics{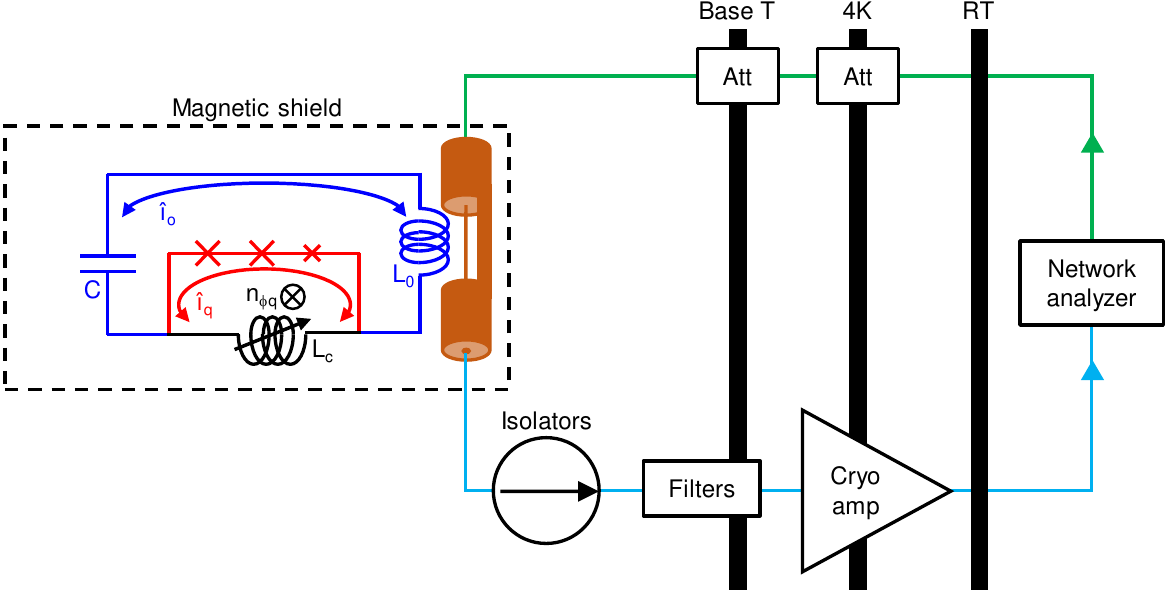}
\caption{\textbf{Measurement setup.} 
The sample of flux qubit coupled to the LC oscillator is cooled down in a dilution refrigerator and measured using a network analyzer. 
For the sample details, see Fig.~1 in the main text.
The green and cyan lines are the signal input and output lines, respectively.
\label{setup}
}
\end{figure}

\begin{figure}
\includegraphics{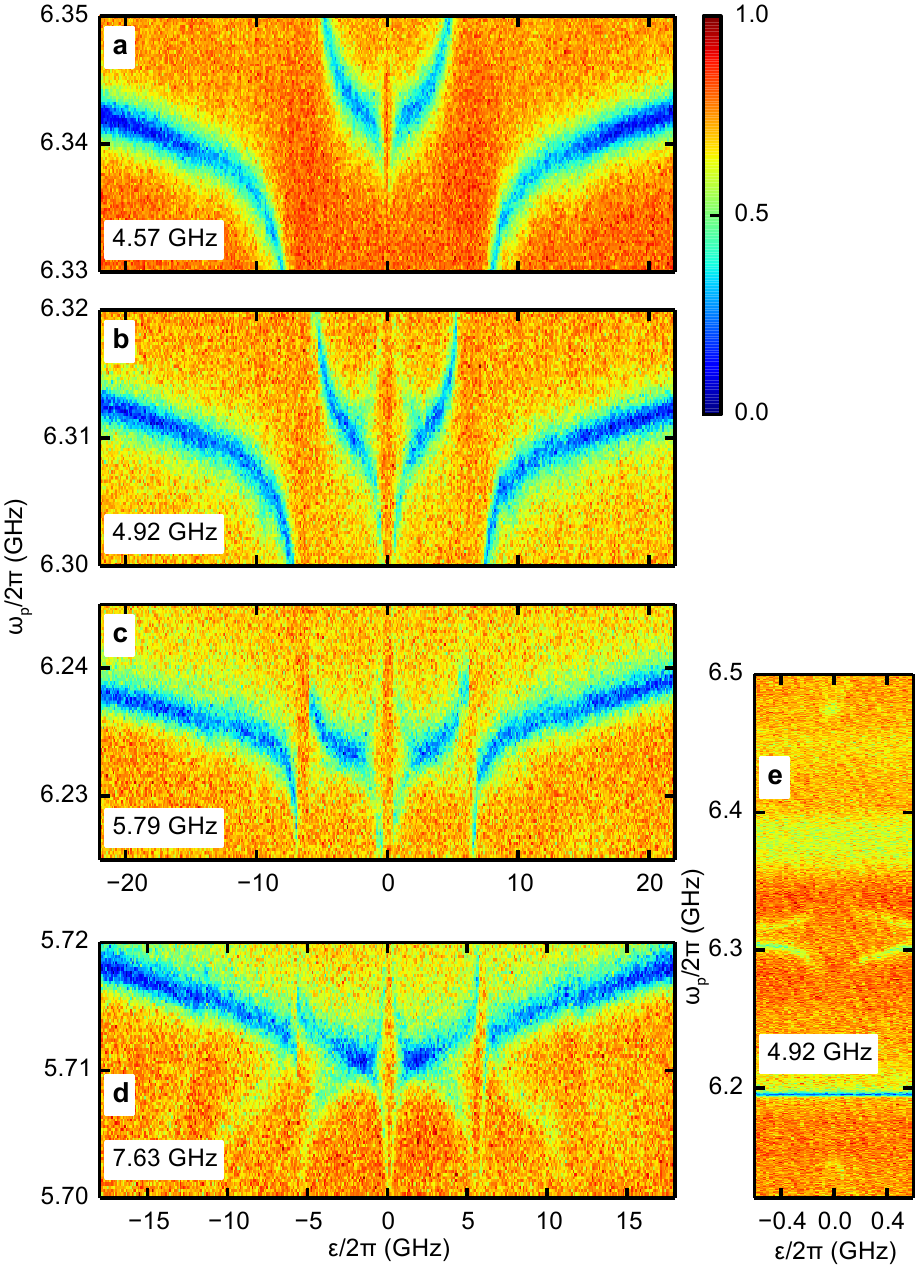}
\caption{\textbf{Transmission spectra for circuit~I and II without the fitting curves.}
The normalized amplitude of the transmission spectra as functions of flux bias $\varepsilon$.
These spectra are the same as Fig.~2\textbf{a}--\textbf{d} and 3\textbf{a} in the main text without calculated transition frequencies $\omega_{ij}^{\rm cal}$.
As summarized in Table~I in the main text, panel \textbf{a} shows data from circuit~I at $n_{\phi \rm q} = -0.5$,
panels \textbf{b} and \textbf{e} show data from circuit~I at $n_{\phi \rm q} = -1.5$,
panel \textbf{c} shows data from circuit~I at $n_{\phi \rm q} = 2.5$,
and panel \textbf{d} shows data from circuit~II at $n_{\phi \rm q} = -0.5$.
The values of $g/2\pi$ are written in the panels.
\label{sparaB4}
}
\end{figure}

\begin{figure}
\includegraphics{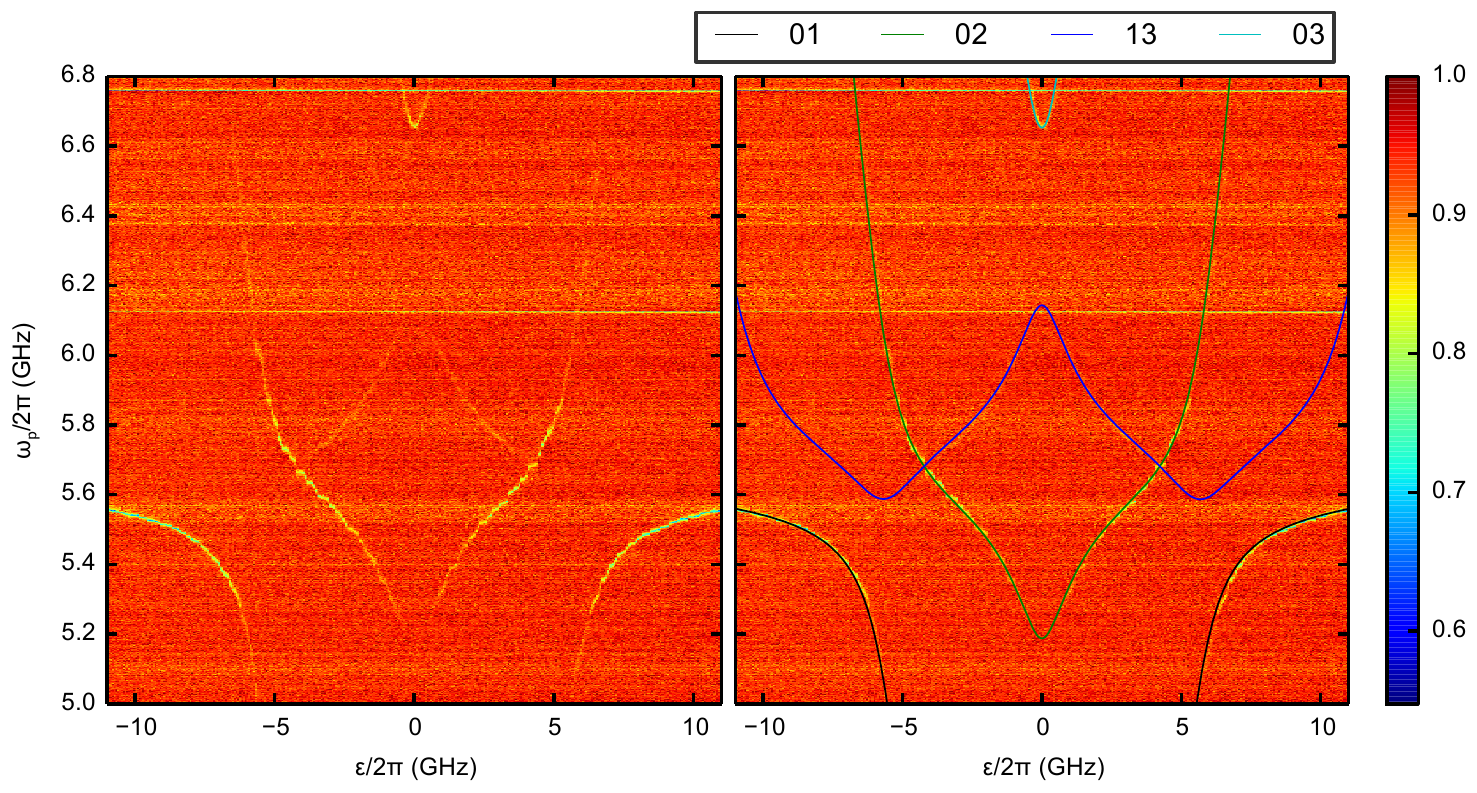}
\caption{\textbf{Transmission spectra for circuit~III.}
The normalized amplitude of the transmission spectra (with calculated transition frequencies $\omega_{ij}^{\rm cal}$ in the right panel).
The black, green, blue, and cyan lines indicate the transitions $|0\rangle \to |1\rangle$, $|0\rangle \to |2\rangle$, $|1\rangle \to |3\rangle$, and $|0\rangle \to |3\rangle$, respectively.
The horizontal signals at 6.12~GHz and 6.75~GHz are from other qubit-oscillator circuits that are coupled to the transmission line and can be ignored here.
\label{sparaBIII}
}
\end{figure}

On each of two sample chips that we prepared, there are four qubit-oscillator circuits
coupled to a single coplanar transmission line.
In order to make them easily identifiable,
we designed the four oscillators to have different resonant frequencies and the four qubits to have different areas.
The energy spectroscopy of the qubit-oscillator circuit is performed via the coplanar transmission line, which is inductively coupled to the LC oscillator, as shown in Fig.~\ref{setup}.
The probe microwave signal is continuous, sent from a network analyzer (Agilent N5234A), and attenuated in the signal input line before arriving at the sample, which is placed in a magnetic shield. 
The transmitted signal from the sample is amplified (by Caltech cryogenic LNA model CITCRYO1-12A) and measured by the network analyzer. 
When the frequency of the probe signal $\omega_{\rm p}$ matches the frequency of a transition between two energy levels, the transmission amplitude decreases, provided that the transition matrix element is not zero.
The input power is kept as low as possible to avoid cascade transitions, such as the transition $|i\rangle \to |j\rangle $ followed by $|j\rangle \to |k\rangle$ when $\omega_{\rm p} \simeq \omega_{ij} \simeq \omega_{jk}$.
The samples are measured in a dilution refrigerator with a nominal base temperature of 10~mK.
From the depth ratio of the signals from the $|0\rangle \to |2\rangle$ and $|1\rangle \to |3\rangle$ transitions shown in Fig.~3\textbf{a} in the main text,
which is directly related to the population ratio of the states $|0\rangle$ and $|1\rangle$,
the temperature of circuit~I at $n_{\phi}=-1.5$ can be estimated to be approximately 45~mK.
In Figs.~2 and 3 in the main text and Figs.~\ref{sparaB4} and \ref{sparaBIII}, the transmission spectrum $S_{21}(\omega_{\rm p})$ is measured at each flux bias $\varepsilon $, and $|S_{21}(\omega_{\rm p}, \varepsilon )|$ is shown.
%The signal at 6.2~GHz in panel \textbf{f} (\textbf{g}) is from another qubit-oscillator circuit that is coupled to the same transmission line but is not relevant to the circuits considered in the main text.

\section{Wigner function of the reduced density operators of the oscillator}
\begin{figure}
\includegraphics{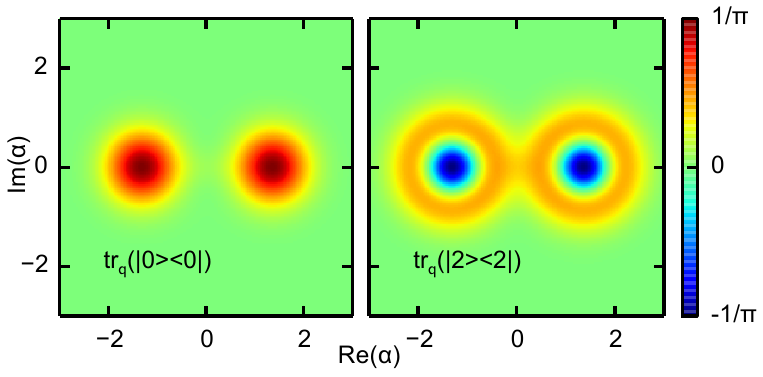}
\caption{\textbf{The calculated Wigner functions of the reduced density operators of the oscillator.}
%The Wigner function of the density operator $\rho$ is given as
%$W(\alpha, \rho)=(1/\pi^2)\int\exp(\eta^*\alpha-\eta\alpha^*)\textrm{tr}[\rho \exp(\eta \hat{a}^{\dagger}-\eta^*\hat{a})]d^2\eta$.
The left and right hand side of the panels shows the Wigner functions of the reduced density operators of the oscillator tr$_{\rm q}(|0\rangle \langle0|)$ and tr$_{\rm q}(|2\rangle \langle 2|)$, respectively,
in the case of circuit II.
The states $|0\rangle$ and
$|2\rangle$ are calculated from $\mathcal{H}_{\rm  total}$ using the parameters in Table I in the main text, and tr$_{\rm q}$ is the partial trace over
qubit states.
\label{Wigs}
}
\end{figure}

Fig.~\ref{Wigs} shows the Wigner functions~\cite{walls2007quantum}, $W(\alpha, \rho)=(1/\pi^2)\int\exp(\eta^*\alpha-\eta\alpha^*)\textrm{tr}[\rho \exp(\eta \hat{a}^{\dagger}-\eta^*\hat{a})]d^2\eta$, of the reduced density operators of the oscillator tr$_{\rm q}(|0\rangle \langle0|)$ and tr$_{\rm q}(|2\rangle \langle 2|)$ in the case of circuit~II,
where
the states $|0\rangle$ and $|2\rangle$ are calculated from $\mathcal{H}_{\rm  total}$ using the parameters in Table~I in the main text, and
tr$_{\rm q}$ is the partial trace over qubit states.
The states tr$_{\rm q}(|0\rangle \langle0|)$ and tr$_{\rm q}(|2\rangle \langle2|)$ are well described by mixtures of the two coherent states $|\pm\alpha\rangle$ and the two displaced Fock states $\hat{D}(\pm \alpha)|1\rangle$ separated from each other by $2\alpha= 2.67$,
where the overlap between the two coherent states is $\langle -\alpha | \alpha \rangle$ = 0.028.

\section{Evaluation of qubit-oscillator entanglement}
\begin{figure}
\includegraphics{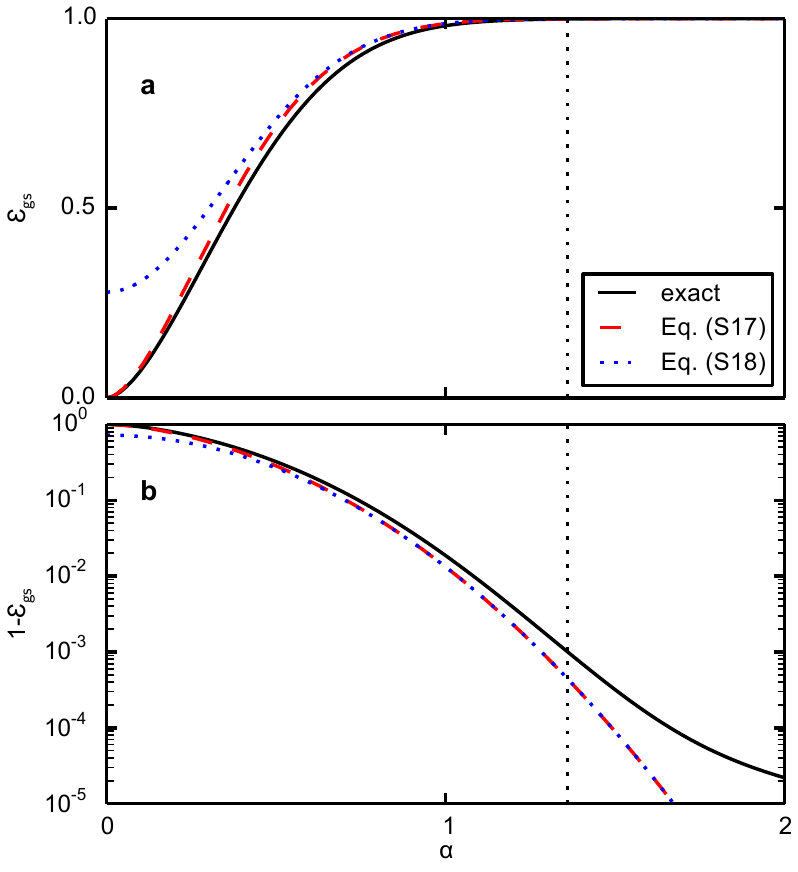}
\caption{\textbf{The qubit-oscillator entanglement as a function of $\alpha$.}
The entanglement is calculated from Eq.~(\ref{Eq:ApproximateExpressionForEntanglement}) (red dashed line), Eq.~(\ref{Eq:FurtherApproximatedExpressionForEntanglement}) (blue dotted line),
and a numerical diagonalization of the Rabi model Hamiltonian using the parameters $\Delta$ and $\omega_{\rm o}$ from circuit II in our experiment (black solid line).
Panel~\textbf{a} shows the linear plot of $\mathcal{E}_{\rm gs}$ while panel~\textbf{b} shows the log plot of $1-\mathcal{E}_{\rm gs}$.
The black dotted line indicates the displacement of circuit~II, i.e. $\alpha = 1.36$.
Both Eq.~(\ref{Eq:ApproximateExpressionForEntanglement}) and Eq.~(\ref{Eq:FurtherApproximatedExpressionForEntanglement}) slightly overestimate the entanglement when $\alpha>1.5$, where the entanglement is around 99.9\% or higher.
Since Eq.~(\ref{Eq:FurtherApproximatedExpressionForEntanglement}) is derived by assuming a large value of $\alpha$, it is not valid when $\alpha<0.5$.
\label{qoent}
}
\end{figure}

The qubit-oscillator entanglement in the ground state can be evaluated as the (base-2) von Neumann entropy of the qubit~\cite{Horodecki09RMP}:
\begin{equation}
\label{Eq:vNEntanglement}
\mathcal{E}_{\rm gs}  =  - {\rm Tr} \{ \rho_{\rm q} \log_2 \rho_{\rm q} \},
\end{equation}
where $\rho_{\rm q}$ is the qubit's reduced density matrix obtained by tracing out the oscillator degree of freedom from the qubit-oscillator ground state.
Figure~\ref{qoent} shows $\mathcal{E}_{\rm gs}$ as a function of $\alpha$,
where $\alpha$ is $g/\omega_{\rm o}$.
We can see from this figure that $\mathcal{E}_{\rm gs}$ increases and approaches 1
 as $\alpha$ increases above 1.

The entanglement for the thermal-equilibrium state can be evaluated as twice the Negativity~\cite{Horodecki09RMP} (which is one of the mixed-state entanglement measures used in the literature):
\begin{eqnarray}
\mathcal{E}_{\rm te} = 2\mathcal{N} = 2\sum_{\lambda < 0}|\lambda|,
\end{eqnarray}
where $\lambda$ are eigenvalues of $\rho^{\Gamma}$,
$\rho$ is the thermal-equilibrium density matrix of the qubit-oscillator system, the superscript $\Gamma$ indicates taking the partial transpose with respect to the degree of freedom of either the qubit or the oscillator,
and the sum is taken over the negative eigenvalues only.
The factor 2 is used to make $\mathcal{E}_{\rm te}$ range from 0 to 1.

\begin{figure}
\includegraphics{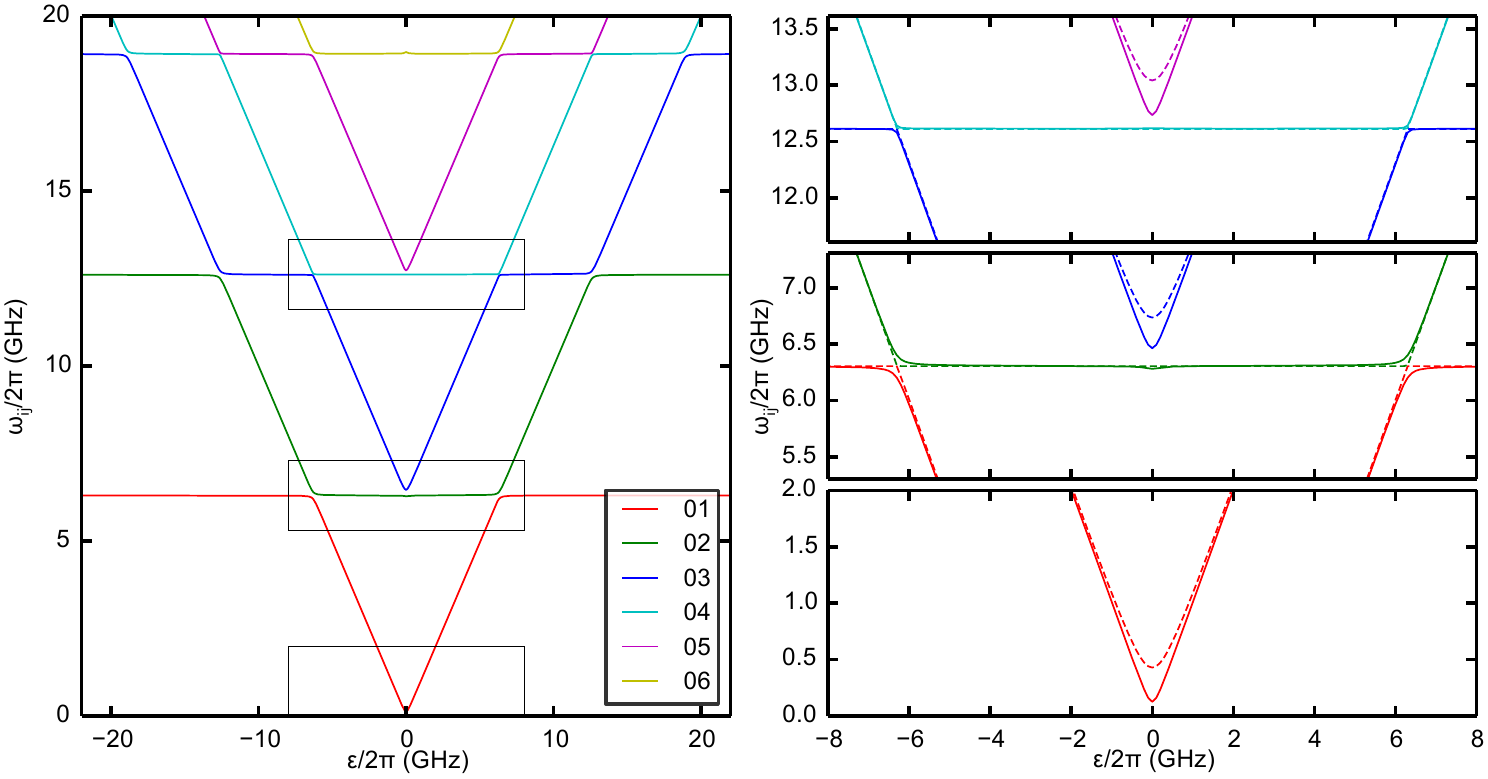}
\caption{\textbf{The calculated transition frequencies from the ground state $\omega_{0i}^{\rm cal}$ as functions of $\varepsilon$.}
The parameters of the circuit I at $n_{\phi \rm q} = -1.5$, i.e., $\Delta/2\pi = 0.430$~GHz, $\omega_{\rm o}/2\pi = 6.306$~GHz, and $g/2\pi = 4.92$~GHz, are used for the calculation.
The three right panels,
where $\omega_{0i}^{\rm cal}$ in the non-interacting case $g = 0$ are also plotted in dashed lines,
are the magnifications of the three rectangles in the left panel.
\label{w0i}
}
\end{figure}

Numerically calculated values for the ground-state entanglements $\mathcal{E}_{\rm gs}$ and thermal-equilibrium entanglements $\mathcal{E}_{\rm te}$ for all five sets of spectroscopy data in the three circuits are summarized in Table~\ref{entT}.
In all cases, $\mathcal{E}_{\rm gs}$ is quite high while $\mathcal{E}_{\rm te}$ is substantially lower than $\mathcal{E}_{\rm gs}$ due to significant population in the state $|1\rangle$.
For circuits~I and II, which have similar values of $\Delta$, $\mathcal{E}_{\rm te}$ decreases as $g$ increases,
which is explained by the suppression of the qubit frequency at $\varepsilon = 0$ from its bare value (see Fig.~\ref{w0i} and Ref.~\cite{Ashhab10PRA}):
\begin{equation}
\label{w01}
\omega_{01}=\Delta e^{-2\alpha^2},
\end{equation}
and the resulting increase in thermal excitation of the state $|1\rangle$.
The relatively high value of  $\mathcal{E}_{\rm te}$ for circuit~III is due to its large $\Delta$ and hence lower population in the state $|1\rangle$.
We have calculated $\mathcal{E}_{\rm gs}$ for the parameters of Refs.~\cite{Niemczyk10} and \cite{FornDiaz10} and found that the values are, respectively, 6\% and 4\%.
Note that in calculating $\mathcal{E}_{\rm gs}$ for Ref.~\cite{Niemczyk10}, out of the different modes of the coplanar waveguide resonator we only consider the mode that is most strongly coupled to the qubit.
If we include all the three modes that are relevant to that experiment, we find that the entanglement between the qubit and the three harmonic oscillators combined is 11\%.

\renewcommand{\figurename}{Table}
\setcounter{figure}{0}
\begin{figure}
\caption{\textbf{Parameters obtained from fitting spectroscopy measurements and calculated entanglement.}}
\includegraphics{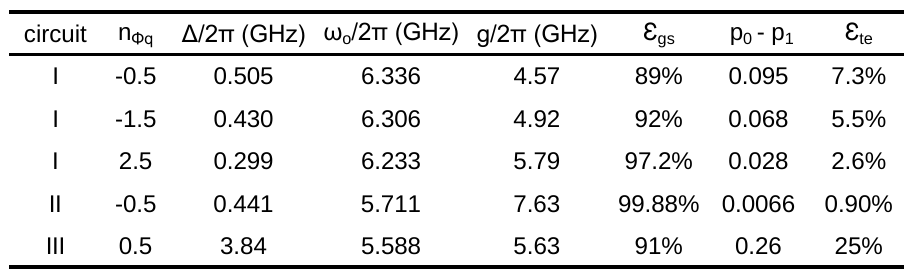}
\label{entT}
\footnote[0]{The parameters are obtained from five sets of spectroscopy data in three circuits.
The ground-state qubit-oscillator entanglement $\mathcal{E}_{\rm gs}$ and the thermal-equilibrium entanglement $\mathcal{E}_{\rm te}$ are calculated numerically using the energies and energy eigenstates obtained using Hamiltonian diagonalization and are essentially exact.
In the calculations of $p_0 - p_1$ and $\mathcal{E}_{\rm te}$, $T = 45$~mK is used.}
\end{figure}

Although $\mathcal{E}_{\rm gs}$ and $\mathcal{E}_{\rm te}$ can be evaluated by numerical calculation,
it is worth deriving approximate expressions for them, which gives a more intuitive picture of the qubit-oscillator entanglement.
The qubit-oscillator ground state at $\varepsilon = 0$ can be approximated as
\begin{equation}
\ket{0} = \frac{1}{\sqrt{2}} \left( \ket{\rm L}_{\rm q} \otimes \ket{-\alpha}_{\rm o} + \ket{\rm R}_{\rm q} \otimes \ket{\alpha}_{\rm o} \right).
\label{Eq:ApproximateGroundState}
\end{equation}
Taking into consideration the fact that $\langle \alpha | -\alpha \rangle = e^{-2\alpha^2}$, the qubit's reduced density matrix is given by
\begin{equation}
\rho_{\rm q} = \frac{1}{2} \left( \begin{array}{cc}
1 & e^{-2\alpha^2} \\
e^{-2\alpha^2} & 1
\end{array} \right).
\end{equation}
The eigenvalues of $\rho_{\rm q}$ are then $\left(1\pm e^{-2\alpha^2}\right)/2$.
%The von Neumann entropy of the qubit is therefore given by
The entanglement can be evaluated using Eq.~(\ref{Eq:vNEntanglement}):
\begin{widetext}
\begin{eqnarray}
\label{Eq:ApproximateExpressionForEntanglement}
\mathcal{E}_{\rm gs}& = & - \frac{1}{2} \left( 1 + e^{-2\alpha^2} \right) \log_2 \left( \frac{1+ e^{-2\alpha^2}}{2} \right) - \frac{1}{2} \left(1- e^{-2\alpha^2} \right) \log_2 \left( \frac{1- e^{-2\alpha^2}}{2} \right),
\end{eqnarray}
\end{widetext}
which when expanded to second order in $e^{-2\alpha^2}$ gives
\begin{equation}
\mathcal{E}_{\rm gs} \simeq 1-\frac{1}{2\ln 2} e^{-4\alpha^2}.
\label{Eq:FurtherApproximatedExpressionForEntanglement}
\end{equation}

Figure~\ref{qoent} shows the entanglement calculated based on the approximate expression for the ground state Eq.~(\ref{Eq:ApproximateGroundState}), with and without the small $e^{-2\alpha^2}$ approximation, along with the entanglement obtained for the numerically calculated (and essentially exact) ground state. For the parameters of circuit II, Eqs.~(\ref{Eq:ApproximateExpressionForEntanglement}) and (\ref{Eq:FurtherApproximatedExpressionForEntanglement}) give an entanglement of 99.94\% while the exact calculation gives the value 99.88\%.

Equation (\ref{Eq:FurtherApproximatedExpressionForEntanglement}) is a poor approximation for $\alpha<0.5$, because when we expanded the logarithm in a Taylor series we assumed a small value of $e^{-2\alpha^2}$.
Figure~\ref{qoent} suggests that Eqs.~(\ref{Eq:ApproximateExpressionForEntanglement}) and (\ref{Eq:FurtherApproximatedExpressionForEntanglement}) deviate from the exact result for $\alpha>1.5$ as well. It should be noted, however, that the absolute value of the error in these approximate expressions decreases monotonically and approaches zero as $\alpha\rightarrow\infty$. It is only when the error is compared to the rapidly decreasing quantity $1-\mathcal{E}_{\rm gs}$ that the approximate expressions seem to deviate from the exact result for large values of $\alpha$.

It can also be seen in Fig.~\ref{qoent} that the approximate expression Eq.~(\ref{Eq:ApproximateGroundState}) leads to an overestimation of the entanglement. This overestimation is due to the fact that Eq.~(\ref{Eq:ApproximateGroundState}) is obtained by ignoring the $\sigma_x$ term in the Hamiltonian (except for its role in identifying the symmetric superposition as the ground state of the coupled system). Because this term does not contain oscillator operators, it favours having a superposition of the states $\ket{\rm{L}}_{\rm q}$ and $\ket{\rm{R}}_{\rm q}$ with the state of the oscillator being independent of the state of the qubit. It therefore favours a slightly increased overlap (in the state of the oscillator) between the two branches of the superposition than the overlap present in Eq.~(\ref{Eq:ApproximateGroundState}), and the increased overlap leads to a reduction in the entanglement. 

A quick estimate for the entanglement in the thermal-equilibrium state $\mathcal{E}_{\rm te}$ can be obtained by taking the product
\begin{equation}
\label{Ete_approx}
\mathcal{E}_{\rm te} \simeq \mathcal{E}_{\rm gs} \times (p_0-p_1) = \left( 1 - \frac{e^{-4\alpha^2}}{2 \ln 2} \right) \times \tanh\left(\frac{\hbar\Delta e^{-2\alpha^2}}{2k_{\rm B}T}\right),
\end{equation}
where $p_0$ and $p_1$ are, respectively, the occupation probabilities of the states $|0\rangle$ and $|1\rangle$,
and $k_{\rm B}$ is the Boltzmann constant.
%
%For small values of $\alpha$, the above quantity is almost equal to $S$ but quickly drops back to zero when $\alpha$ increases beyond the point where $\hbar\omega_{01}$ exceeds $k_BT$.
%When $\alpha$ increases, $\mathcal{E}_{\rm gs}$ approaches to one while $p_0 - p_1$ approaches to zero.
When $\alpha=0$, one obviously has $\mathcal{E}_{\rm te}=\mathcal{E}_{\rm gs}=0$ (although the right-most part of Eq.~(\ref{Ete_approx}) gives a finite value because it contains a poor approximation for $\mathcal{E}_{\rm gs}$ in that limit).
On the other hand, when $\alpha$ increases to very large values, $\mathcal{E}_{\rm gs}$ approaches one but $p_0-p_1$ approaches zero.
As a result, there is an optimal value of $\alpha$ that balances between maximizing the ground-state entanglement and maximizing the ground-state occupation probability.

Although the above estimate for the thermal-equilibrium entanglement might seem very hand-waving,
it turns out to be a rather good estimate, especially in the limits of large $\mathcal{E}_{\rm gs}$ or large $p_0-p_1$.
For example, if we consider a statistical mixture of two complementary two-qubit Bell states,
e.g.~$(\ket{1}\otimes\ket{0}\pm\ket{0}\otimes\ket{1})/\sqrt{2}$, with probabilities $p_0$ and $p_1$,
the negativity multiplied by two takes the simple form
\begin{equation} 
2 \mathcal{N} = | p_0 - p_1 |.
\end{equation}
As a result, our estimate is rather accurate when $\alpha$ is large and therefore $\mathcal{E}_{\rm gs}$ is very close to one,
because in this limit the overlap between the oscillator states $\ket{\alpha}$ and $\ket{-\alpha}$ approaches zero and the lowest two energy eigenstates do indeed form a pair of complementary Bell states.
Similarly, if we take the opposite limit where $\hbar\omega_{01}\gg k_{\rm B}T$ and $p_0-p_1$ is close to one,
thermal population of the excited states can be ignored,
and the ground-state entanglement $\mathcal{E}_{\rm gs}$ becomes a good estimate for the thermal-equilibrium entanglement.

Before concluding, it is worth pointing out here the difference between the coefficient 4 inside the exponent in Eq.~(\ref{Eq:FurtherApproximatedExpressionForEntanglement}) and the coefficient 2 inside the exponent in Eq.~(\ref{w01}).
The entanglement therefore approaches 100\% much faster than $\omega_{01}$ approaches zero.
This property is desirable for future designs to achieve a high thermal-equilibrium entanglement.

\end{document}